\documentclass[aps,prd,showpacs,notitlepage,nofootinbib,preprintnumbers,amsmath,amssymb]{revtex4-1}

\usepackage{graphics,graphicx}
\usepackage{dcolumn}
\usepackage{bm}
\usepackage{epsfig}
\usepackage[usenames]{color}
\usepackage{hyperref} 
\usepackage{mathbbol}
\usepackage{epstopdf}
\usepackage{simplewick}

\def\eq#1{{Eq.~(\ref{#1})}}
\def\fig#1{{Fig.~\ref{#1}}}
\newcommand{\ben}{\begin{eqnarray*}}
\newcommand{\een}{\end{eqnarray*}}

\newcommand{\pd}{\partial}

\newcommand{\dhd}{{\textstyle d}
\lower.03ex\hbox{\kern-0.38em$^{\scriptstyle-}$}\kern-0.05em{}}
\newcommand{\dbar}{{\textstyle \delta}
\lower.03ex\hbox{\kern-0.38em$^{\scriptstyle-}$}\kern-0.05em{}}
\newcommand{\half}{{1\over 2}}

\newcommand{\cald}{{\cal D}}

\newcommand{\tilden}{\eta}

\newcommand{\teta}{{\tilde\eta}}

\begin{document}

\title{Regularization of the Light-Cone Gauge Gluon Propagator
  Singularities \\ Using Sub-Gauge Conditions}

\author{Giovanni~A.~Chirilli,\footnote{chirilli.1@osu.edu}
  Yuri~V.~Kovchegov\footnote{kovchegov.1@osu.edu},
  Douglas~E.~Wertepny\footnote{wertepny.1@osu.edu}}

\affiliation{Department of Physics, The Ohio State University,
  Columbus, OH 43210, USA}

\begin{abstract}
  Perturbative QCD calculations in the light-cone gauge have long
  suffered from the ambiguity associated with the regularization of
  the poles in the gluon propagator. In this work we study sub-gauge
  conditions within the light-cone gauge corresponding to several
  known ways of regulating the gluon propagator. Using the functional
  integral calculation of the gluon propagator, we rederive the known
  sub-gauge conditions for the $\theta$-function gauges and identify
  the sub-gauge condition for the principal value (PV) regularization
  of the gluon propagator's light-cone poles. The obtained sub-gauge
  condition for the PV case is further verified by a sample
  calculation of the classical Yang-Mills field of two collinear
  ultrarelativistic point color charges. Our method does not allow one
  to construct a sub-gauge condition corresponding to the well-known
  Mandelstam--Leibbrandt prescription for regulating the gluon
  propagator poles.
\end{abstract}

\pacs{12.38.-t, 12.38.Bx, 12.38.Cy}

\maketitle



\section{Introduction}

Consider a gluon (or photon) propagator in the 
\begin{align}
  \label{LCgauge}
  \eta \cdot A = A^+ = 0
\end{align}
light-cone gauge:
\begin{align}
  D^{\mu\nu} (x,y) \equiv \langle 0 | {\rm T} A^\mu(x) A^\nu(y) | 0
  \rangle = \int {d^4 k \over (2 \pi)^4} \, e^{-ik\cdot(x-y)} \,
  \frac{-i}{k^2 +i\epsilon} \, \left[g^{\mu\nu} - {k^\mu\eta^\nu +
    k^\nu\eta^\mu\over k^+}\right].
\label{prop-LCgauge}
\end{align}
(The gluon propagator given by \eq{prop-LCgauge} is diagonal in the
color indices.) Our convention for four-vectors is $v^\mu = (v^+, v^-,
{\vec v}_\perp)$ with $v^\pm = (v^0 \pm v^3)/\sqrt{2}$. The gauge
condition \eqref{LCgauge} and the propagator \eqref{prop-LCgauge} are
defined with the help of a light-like four-vector
\begin{align}
  \label{eq:eta}
  \eta^\mu \equiv ( 0, 1, {\vec 0}_\perp ),
\end{align}
such that
\begin{align}
  \label{eq:eta2}
  \eta^2 = 0, \ \ \ \eta \cdot x = x^+~.
\end{align}

Using the gluon propagator \eqref{prop-LCgauge} in practical
perturbative calculations one invariably faces the need to find a
suitable way of regulating the $k^+=0$ pole. (See
\cite{Bassetto:1991ue,Bassetto:1984dq} for a retrospective of works on
the subject.)  Without such regularization the $k^+$-integral in
\eq{prop-LCgauge} is ill-defined. The singularity of \eq{prop-LCgauge}
at $k^+=0$ appears to be due to incomplete gauge fixing: the $A^+ =0$
light-cone gauge is preserved under any $x^-$-independent gauge
transformation, given by
\begin{align}
  \label{eq:gauge_tr}
  A^\mu (x) \to A^\mu (x) + \partial^\mu \Lambda (x^+, {\vec x}_\perp)
\end{align}
in the Abelian case and by
\begin{align}
  \label{eq:Sgauge_tr}
  A^\mu (x) \to S(x^+, {\vec x}_\perp) A^\mu (x) S^{-1} (x^+, {\vec
    x}_\perp) - \frac{i}{g} \left[ \partial^\mu S (x^+, {\vec
      x}_\perp) \right] \, S^{-1} (x^+, {\vec x}_\perp)
\end{align}
in the non-Abelian case. It is usually assumed that regularization of
the $k^+=0$ pole should follow from further gauge fixing, stemming
from sub-gauge constraints imposed in addition to \eq{LCgauge}.

The most commonly used regularization prescriptions for the $k^+=0$
pole of the gluon light-cone gauge propagator are as follows:
\begin{itemize}
\item $\theta$-function sub-gauges
  \cite{Slavnov:1987yh,Kovchegov:1997pc,Belitsky:2002sm}:
\begin{align}
  D_1^{\mu\nu}(x, y) \equiv \int \frac{d^4k}{(2 \pi)^4} \,
  e^{-ik\cdot(x-y)} \, {-i\over k^2+i\epsilon} \left[ g^{\mu\nu} - {k^\mu
    \eta^\nu \over k^+-i\epsilon} - {k^\nu \eta^\mu \over
    k^++i\epsilon} \right] ,
  \label{propagator-10} \\
  D_2^{\mu\nu}(x, y) \equiv \int \frac{d^4k}{(2 \pi)^4} \,
  e^{-ik\cdot(x-y)} \, {-i\over k^2+i\epsilon} \left[ g^{\mu\nu} - {k^\mu
    \eta^\nu \over k^++i\epsilon} - {k^\nu \eta^\mu \over
    k^+-i\epsilon}\right] .
	\label{propagator-20}
\end{align}
The name stems from the fact that the classical field of a point
(color) charge moving along the $x^- =0$ light cone is proportional to
$A^\mu_\perp \sim \theta (-x^-)$ in the first case and $A^\mu_\perp
\sim \theta (x^-)$ in the second case
\cite{Mueller:1988xy,McLerran:1993ni,McLerran:1993ka,McLerran:1994vd,Kovchegov:1996ty,Kovchegov:1997pc}.

\item Principal value (PV) sub-gauge \cite{Curci:1980uw}
  \begin{align}
    D_{PV}^{\mu\nu}(x, y) \equiv \int \frac{d^4k}{(2 \pi)^4} \,
    e^{-ik\cdot(x-y)} \, {-i\over k^2+i\epsilon} \, \left[ g^{\mu\nu} -
    \Big(k^\mu \eta^\nu + k^\nu \eta^\mu\Big){\rm PV}\Big\{{1\over
      k^+}\Big\}\right] .
\label{propagator-PV0}
  \end{align}

\item Mandelstam--Leibbrandt (ML) prescription
  \cite{Mandelstam:1982cb,Leibbrandt:1983pj}
\begin{align}
  D^{\mu\nu}_{ML}(x, y) = \int \!\! \frac{d^4k}{(2 \pi)^4} \, e^{-i
    k\cdot(x-y)} \, {-i\over k^2+i\epsilon}\left[g^{\mu\nu} -
  {k^\mu\eta^\nu + k^\nu\eta^\mu \over k^+ + i\epsilon k^-}\right] ~.
\label{ML-propag0}
\end{align}

\end{itemize}

The goal of this work is to identify the sub-gauge conditions leading
to the propagators in Eqs.~\eqref{propagator-10},
\eqref{propagator-20}, \eqref{propagator-PV0} and \eqref{ML-propag0}
and to demonstrate that these sub-gauge conditions result in the
propagators listed in those formulas when implemented in Feynman
functional integration. We would like to stress that the
regularizations of the gluon propagator poles given in
Eqs.~\eqref{propagator-10}, \eqref{propagator-20},
\eqref{propagator-PV0} and \eqref{ML-propag0} are by no means
exhaustive, and other regularizations exist which will not be
considered in this work (see e.g. \cite{Das:2004qk}).

The paper is structured as follows. We begin with the
$\theta$-function sub-gauges in Sec.~\ref{sec:theta_ftn}. Motivated by
the $A^0=0$ gauge we propose the sub-gauge condition in \eq{subgauge},
impose this sub-gauge condition within the functional integral, and
derive an expression for the gluon propagator (with the $k^+ =0$ pole
regulated) by carefully evaluating surface terms inside the functional
integral. In the process we show that the sub-gauge condition
\eqref{subgauge} can only be imposed at $x^- = \pm \infty$. The final
results for the light-cone gluon propagators are given in
Eqs.~\eqref{propagator-1} and \eqref{propagator-2}, with the
corresponding sub-gauge conditions stated immediately above these
propagators. The same sub-gauge conditions were employed previously in
\cite{Belitsky:2002sm,Kovner:1995ts}.

We move on to the case of the PV sub-gauge in Sec.~\ref{sec:PV}. There
we tackle the problem in reverse order: we search for a sub-gauge
condition which yields the propagator \eqref{propagator-PV0} in the
functional integral calculation similar to that in
Sec.~\ref{sec:theta_ftn}. In the end we obtain
\begin{align}
  {\vec \partial}_\perp\cdot {\vec A}_\perp(x^-=+\infty) +
  {\vec \partial}_\perp\cdot {\vec A}_\perp(x^-=-\infty) =0
\label{PV-subgauge0}
\end{align}
as the sub-gauge condition necessary to obtain the PV regularization
of the light-cone gauge gluon propagator \eqref{propagator-PV0}.

The same reverse strategy is applied to the Mandelstam--Leibbrandt
prescription in Sec.~\ref{sec:ML}. Starting from the
Mandelstam--Leibbrandt propagator \eqref{ML-propag0} we try to
reconstruct the sub-gauge condition corresponding to this
propagator. Unfortunately this procedure fails to yield a valid
sub-gauge condition for the ML case.

Finally, in Sec.~\ref{Class} we illustrate and test our conclusion
about the proper sub-gauge fixing condition \eqref{PV-subgauge0} for
the PV case by constructing the classical gluon field of two
ultrarelativistic color charges moving in the same direction.
Problems like this arise in describing the gluon distribution of a
single large nucleus in the framework of the McLerran--Venugopalan
(MV) model
\cite{McLerran:1993ni,McLerran:1993ka,McLerran:1994vd,Kovchegov:1996ty,Kovchegov:1997pc,Jalilian-Marian:1997xn}. The
classical gluon field of a single nucleus was constructed in the MV
model in \cite{Kovchegov:1996ty,Jalilian-Marian:1997xn} by using one
of the $\theta$-function sub-gauges. In Sec.~\ref{Class}, for the
first time the field is obtained in the PV sub-gauge. The gluon field
is constructed both by solving the classical Yang-Mills equations and
by diagram summation. In particular we show that at the Abelian
lowest-order in the coupling $g$ level one may use the sub-gauge
condition (see e.g. \cite{Belitsky:2002sm})
\begin{align}
  {\vec A}_\perp(x^-=+\infty) + {\vec A}_\perp(x^-=-\infty) =0
\label{PV-subgauge_off}
\end{align}
instead of that in \eq{PV-subgauge0}. However, at higher orders in
$g$, when the non-Abelian corrections become important, it is
impossible to enforce the condition \eqref{PV-subgauge_off} even for
the classical gluon field. At the same time the condition
\eqref{PV-subgauge0} appears to work even at the non-Abelian
level. Combined with the derivation in Sec.~\ref{sec:PV}, this result
appears to put on a more solid footing the PV regularization of
light-cone gluon propagator singularities, which was used in
perturbative calculations in the past \cite{Curci:1980uw}.

We conclude in Sec.~\ref{sec:conclusions} by restating our main
results.


For future use let us define another light-like four-vector,
\begin{align}
  \label{eq:eta}
  {\tilde \eta}^\mu \equiv ( 1, 0, {\vec 0}_\perp ), \ \ \ \teta^2 =0,
  \ \ \ \teta \cdot x = x^-~.
\end{align}
Any four-vector can be decomposed as $k^\mu = k^+ \teta^\mu + k^-
\eta^\mu + k^\mu_\perp$, where $k^\mu_\perp = (0, 0, k^1,k^2)$ and
$a_\perp\cdot b_\perp = a^i b^i = - a_{\perp\mu}b^\mu_\perp$ with
$i=1,2$ and $\mu = 0, \ldots, 3$. We also define ${\vec k}_\perp
\equiv (k^1, k^2)$.


\section{$\theta$-Function Sub-Gauges}

\label{sec:theta_ftn}

In this section we will re-derive the sub-gauge conditions and the
gluon propagator for the $\theta$-function sub-gauges of the $A^+ =0$
light-cone gauge using the functional integral formalism. We start
with a conjecture for the sub-gauge condition. Note that in the case
of temporal $A^0=0$ gauge one has a similar situation: the gluon
propagator and the prescription for regulating the singularity at $k^0
=0$ in it are obtained by imposing a sub-gauge condition at a specific
point in time: $\vec{\partial}\cdot \vec{A}(t_0, \vec{x}) =0$
\cite{Rossi:1979jf,Rossi:1980pg,Leroy:1986uc,Slavnov:1986km,Chirilli}. Motivated
by the $A^0=0$ gauge example, we impose the following sub-gauge
condition:
\begin{align}
  \partial_{\perp\mu}A_\perp^\mu (x^+, x^-=\sigma, {\vec x}_\perp) =
  0~.
\label{subgauge}
\end{align}
In other words, we require that the transverse divergence of the gauge
field vanishes at $x^-=\sigma$ with the value of $\sigma$ not
specified yet. (In the $A^0=0$ gauge the corresponding time $t_0$ at
which the sub-gauge condition is specified remains arbitrary.)
Clearly, \eq{subgauge} is not the only sub-gauge choice that can be
made. For example, an alternative gauge choice is to require that the
four-divergence is zero at a generic point in $x^-$,
$\partial_{\mu}A^\mu (x^+, x^-=\sigma, {\vec x}_\perp) = 0$. However,
as we will explain below (see e.g. Appendix~\ref{B}), this sub-gauge
choice is not supported by the functional integral calculation.

In the functional integral formalism the propagator is obtained
by applying functional derivatives of the generating functional with
respect to the sources,
\begin{align}
  \langle 0 | {\rm T} A_\mu(x) A_\nu(y) | 0 \rangle & = - \left[
    {\delta\over \delta J^\mu(x)}{\delta\over \delta J^\nu(y)}\,
    e^{-\half\int d^4 x' d^4 y'\, J^\alpha (x') D_{\alpha\beta}(x',y')\,
      J^\beta(y')} \right] \Bigg|_{J =0} \notag \\ & = - \left[
    {\delta\over \delta J^\mu(x)}{\delta\over \delta J^\nu(y)}\,
    \left( \frac{Z[J]}{Z[0]} \right) \right] \Bigg|_{J =0} ,
\label{def-prop}
\end{align}
where $D_{\mu\nu}(x,y)$ is the gluon propagator and $Z[J]$ is the
generating functional. To arrive at the expression for the gluon
propagator $D_{\mu\nu}(x,y)$ (with regularizations for all the poles
in momentum space) using the functional integration for constructing
the generating functional used in (\ref{def-prop}), one has to take
special care of the surface terms arising from integration by parts
and of the gauge conditions. In what follows we will consider the
$x^+$ variable as time, and will define the initial and final
conditions at the light-cone times $x_i^+$ and $x_f^+$
respectively. It will be implied that $x_i^+$ is large and negative
while $x_f^+$ is large and positive. In addition we assume that the
system is localized in space but not in time: since now $x^+$ is our
time variable, instead of the ``standard'' assumption that all fields
go to zero as $|{\vec x}| \to \infty$, we will assume that the fields
go to zero as $|{\vec x}_\perp| \to \infty$. As will become apparent
below, careful treatment will be needed of the functional integral at
the boundaries in $x^+$ and $x^-$ directions.

The generating functional for an Abelian gauge theory in the
light-cone gauge with the sub-gauge condition (\ref{subgauge}) is
\begin{align}
  Z[J] = \lim_{\xi_1,\xi_2 \to 0}\int \!\cald A_i \, \cald A_f \Psi_0
  (A_i) \Psi_0^*(A_f) \hspace*{-10mm} \int\limits_{~~~~\substack{
      A(x^+_i, x^-, \vec{x}_\perp)=A_i \\
      A(x^+_f, x^-, \vec{x}_\perp)=A_f }} \hspace*{-12mm} \cald A_\mu
  \, {\rm exp}\left\{i \int^{x^+_f}_{x^+_i} \!\! dx^+ \int \!\! dx^-
    \, d^2 x_\perp \Big[ {\cal L}_0 (A) + {\cal L}_{fix} (A) + J_\mu
    A^\mu\Big]\right\}
\label{fully-fixed-func}
\end{align}
with
\begin{align} 
  {\cal L}_0 (A) = - \frac{1}{4} \, F_{\mu\nu} \, F^{\mu\nu} = -\half
  (\partial_\mu A_\nu)(\partial^\mu A^\nu) + \half (\partial_\mu
  A_\nu)(\partial^\nu A^\mu)
\end{align}
and the gauge and sub-gauge fixing terms 
\begin{align} 
  {\cal L}_{fix} (A) = - {1\over 2\xi_1} \, A_\mu \, \eta^\mu \,
  \eta^\nu \, A_\nu -{1\over 2\xi_2} \, \left( {\vec \partial}_\perp
    \cdot {\vec A}_\perp \right)^2 \, \delta(x^- -\sigma).
\end{align}
The generating functional in \eq{fully-fixed-func} can also be thought
of as describing the Abelian part of a non-Abelian theory such as
gluodynamics. Notice that, as discussed above, in the generating
functional (\ref{fully-fixed-func}) we have used the light-cone
coordinates with $x^+$ as the time direction. As is usually done, we
have exponentiated the gauge conditions and the parameters $\xi_1$ and
$\xi_2$ will be sent to zero at the end of the calculation.

In \eq{fully-fixed-func} $\Psi_0 (A)$ represents the vacuum wave
function in the $A_\mu$-representation.  In the light-cone gauge it is
\begin{align}
  \label{vac_wf}
  \Psi_0 (A) = {\rm exp}\left\{\half\int d x^- d^2x_\perp \,
  A^\mu\sqrt{-(\partial^+)^2}A_\mu \right\} .
\end{align}
The expression \eqref{vac_wf} can be obtained by starting with the
vacuum wave function in the $A^0 =0$ gauge (see Eq.~(7.7) in
\cite{Rossi:1979jf})
\begin{align}
  \label{eq:A0wf}
  \Psi_0 (A) = {\rm exp}\left\{ - \half\int d^3x \, A^i \sqrt{- {\vec
        \nabla}^2} \, \left[ \delta^{ij} - \frac{\partial^i
        \, \partial^j}{{\vec \nabla}^2} \right] \, A^j \right\},
\end{align}
(with ${\vec \nabla} = (\pd_x, \pd_y, \pd_z)$ and $i,j = 1,2,3$ only
in this formula) and performing an ultra-boost along the $+z$
direction to change the gauge into the $A^+=0$ gauge and the wave
function \eqref{eq:A0wf} into \eqref{vac_wf}.

It is known that one of the advantages of using axial-type gauge
conditions is the absence of ghost fields. However, now, in addition
to the light-cone gauge, we have a sub-gauge condition
\eqref{subgauge} which introduces a non trivial determinant, leading
to a ghost field $c (x)$ localized at $x^- = \sigma$:
\begin{align}
  \det \left[ \pd^\perp_\mu \, {\cal D}^\mu_\perp (x^- = \sigma)
  \right] = \int \cald {\bar c} \, \cald c \, {\rm exp}\left\{ -
    i \int d x^+ \, d^2 x_\perp \, {\bar c} \, \pd^\perp_\mu \,
    {\cal D}^\mu_\perp \, c (x^- = \sigma) \right\},
\end{align}
where $\cald_\mu^{ab} \equiv \pd_\mu \, \delta^{ab} + g \, f^{acb} \,
A^c_\mu$ is the covariant derivative and ${\bar c} (x)$ is the complex
conjugate ghost field. Just like in Feynman gauge, the ghost field is
needed only in the non-Abelian case. The ghost field does not affect
the gluon propagator in question. The propagator of this ghost field,
along with the ghost-gluon vertices, depend only on transverse
momenta, and are independent of $k^-$. Because of that it appears that
ghost loops are zero in calculations using dimensional
regularization. Therefore, in \eq{fully-fixed-func} and in the
subsequent analysis we omit ghost contributions arising from sub-gauge
conditions.

In order to put \eq{fully-fixed-func} in the same form as the first
line of \eq{def-prop}, we will adopt the following standard procedure
of ``completing the square''. First we perform a shift of the gauge
field $A^\mu \to A^\mu + a^\mu$ and obtain
\begin{align}
  Z[J] = \lim_{\xi_1,\xi_2\to 0}\int & \cald A_i \, \cald A_f \Psi_0
  (A_i) \Psi_0^*(A_f) \, \Psi_0 (a_i) \Psi^*_0 (a_f) \, {\rm
    exp}\left\{ \int d x^- d^2x_\perp \, \Big( A_i^\mu
    \sqrt{-(\partial^+)^2} \, a_{i\,\mu} + A_f^\mu
    \sqrt{-(\partial^+)^2}\, a_{f\,\mu}\Big)\right\}
  \nonumber\\
  \times \int\limits_{~~~~\substack{
      A(x^+_i,x^-, \vec{x}_\perp)=A_i \\
      A(x^+_f,x^-, \vec{x}_\perp)=A_f }} \hspace*{-12mm} \cald A_\mu &
  \ {\rm exp}\Bigg\{ i \int^{x^+_f}_{x^+_i} \!\! dx^+ \int \!\! d x^-
  \, d^2 x_\perp \Bigg[ {\cal L}_0 (A) + {\cal L}_{fix} (A) + {\cal
    L}_0(a) + {\cal L}_{fix}(a) + J^\mu A_\mu + J^\mu a_\mu +
  \nonumber\\
  & - (\partial_\mu A_\nu) \, (\partial^\mu a^\nu) + (\partial_\mu
  A_\nu) \, (\partial^\nu a^\mu) - {1\over \xi_1} A_\mu \, \eta^\mu \,
  \eta^\nu \, a_\nu - {1\over \xi_2}\left( {\vec \partial}_\perp\cdot
    {\vec A}_\perp \right) \, \left({\vec \partial}_\perp\cdot {\vec
      a}_\perp \right) \, \delta(x^- - \sigma)\Bigg]\Bigg\}
  ~. \label{Z1}
\end{align}
In arriving at \eq{Z1} we have done an integration by parts in (parts
of) the vacuum wave functions, discarding the two-dimensional boundary
integral which is outside the precision of the approximation that was
used in deriving \eq{vac_wf}. We now perform integration by parts in
the terms linear in $a^\mu$ in the rest of the expression to arrive at
\begin{align}
  & Z[J] = \lim_{\xi_1,\xi_2\to 0}\int \! \cald A_i \, \cald A_f
  \Psi_0 (A_i) \Psi_0^*(A_f) \, \Psi_0 (a_i) \, \Psi^*_0 (a_f) \, {\rm
    exp}\left\{ \int d x^- d^2x_\perp \, \Big( A_i^\mu
    \sqrt{-(\partial^+)^2} \, a_{i\,\mu} + A_f^\mu
    \sqrt{-(\partial^+)^2}\, a_{f\,\mu}\Big)\right\}
  \nonumber\\
  & \times \int\limits_{~~~~\substack{
      A(x^+_i,x^-, \vec{x}_\perp)=A_i \\
      A(x^+_f,x^-, \vec{x}_\perp)=A_f }} \hspace*{-12mm} \cald A_\mu \
  {\rm exp}\Bigg\{ i \int^{x^+_f}_{x^+_i} \!\! dx^+ \int \!\! d x^- \,
  d^2 x_\perp \Bigg[ {\cal L}_0 (A) + {\cal L}_{fix} (A) + {\cal
    L}_0(a) + {\cal L}_{fix}(a) + J^\mu A_\mu + J^\mu a_\mu +
  \nonumber\\
  & + A_\nu \left[ \partial^2 \, g^{\mu\nu} - \partial^\mu
    \, \partial^\nu \right] a_\mu - {1\over \xi_1} A_\mu \, \eta^\mu
  \, \eta^\nu \, a_\nu + {1\over \xi_2} \,
  A_{\perp\mu}(\partial_\perp^\mu \partial_\perp^\nu a_{\perp\nu})\,
  \delta(x^- - \sigma) \Bigg] - i \int d\sigma_\mu
  \Big[A_\nu(\partial^\mu a^\nu) - A_\nu(\partial^\nu a^\mu)\Big]
  \Bigg\} .
\label{gen-funct}
\end{align}
where $d\sigma^\mu = \pm (d^2x_\perp \, dx^+ \, \teta^\mu + d^2x_\perp
\, d x^- \, \eta^\mu + d\sigma^\mu_\perp)$ is the integration measure
over the 3-dimensional surface of our four-dimensional space-time.
Here $d\sigma^\mu_\perp$ is the integration measure over the surface
at $x_\perp \to \infty$. The choice of a plus or minus in each of the
terms depends on which boundary one is integrating over.

In order to ``complete the square'' we need to eliminate all the terms
linear in $A^\mu$ in \eq{gen-funct}. Starting from the 4-dimensional
volume integration terms we have to choose $a^\mu$ such that
\begin{align}
  A_\nu \left[ \partial^2 \, g^{\mu\nu} - \partial^\mu \, \partial^\nu
  \right] a_\mu - {1\over \xi_1} \, A_\mu \, \eta^\mu \, \eta^\nu \,
  a_\nu + {1\over \xi_2} \,
  A_{\perp\mu}(\partial_\perp^\mu \partial_\perp^\nu a_{\perp\nu}) \,
  \delta(x^- - \sigma) + J_\mu A^\mu = 0
\end{align}
for any $A^\mu$. Solving for $a^\mu$ we get
\begin{align}
  a^\mu (x) = i \, \int d^4 y \, {D}^{\mu\nu}(x,y) \, J_\nu(y)
\label{def-a}
\end{align}
where ${D}^{\mu\nu}(x,y)$ is the Green function found from
\begin{align}
  \left[ \partial^2 g^{\mu\nu} - \partial^\mu\partial^\nu - {1\over
      \xi_1}\tilden^\mu\tilden^\nu + {1\over
      \xi_2}\partial_\perp^\mu\partial_\perp^\nu \, \delta(x^- -
    \sigma)\right] \, {D}_{\nu\rho}(x,y) = i \, \delta_\rho^\mu \,
  \delta^{(4)}(x-y).
\label{greenf-eq}
\end{align}

The boundary conditions for \eq{greenf-eq} are obtained by requiring
that the 3-dimensional {\sl surface} integration terms linear in
$A^\mu$ should also vanish in the exponent of \eq{gen-funct},
\begin{align}
  \int d x^- \, d^2x_\perp \, \Big( A_i^\mu \sqrt{-(\partial^+)^2} \,
  a_{i\,\mu} + A_f^\mu \sqrt{-(\partial^+)^2} \, a_{f\,\mu}\Big) - i
  \int d\sigma_\mu \Big[ A_\nu(\partial^\mu a^\nu) -
  A_\nu(\partial^\nu a^\mu)\Big] =0.
\label{vanish-terms1}
\end{align}
Note that the condition \eqref{vanish-terms1} eliminates all the
boundary term dependent on $a^\mu$ from the exponent of \eq{gen-funct}
(and not just the terms linear in $A^\mu$). More precisely, for
$a^\mu$ satisfying \eqref{vanish-terms1} one gets
\begin{align}
  & \Psi_0 (a_i) \, \Psi_0^* (a_f) \, {\rm exp}\Big\{\int d x^- d^2x
  \Big( A_i^\mu \sqrt{-(\partial^+)^2}a_{i\,\mu} + A_f^\mu
  \sqrt{-(\partial^+)^2}a_{f\,\mu}\Big)\Big\}
  \nonumber\\
  & \times{\rm exp}\Bigg\{ - \frac{i}{2} \int d\sigma_\mu
  \Big(a_\nu(\partial^\mu a^\nu) - a_\nu(\partial^\nu a^\mu)\Big) - i
  \int d\sigma_\mu \Big(A_\nu(\partial^\mu a^\nu) - A_\nu(\partial^\nu
  a^\mu)\Big) \Bigg\} = 1.
 \label{vanish-terms}
\end{align}

With this in mind one can readily show that after using $a^\mu$
satisfying Eqs.~(\ref{def-a}), \eqref{greenf-eq} and
\eqref{vanish-terms1} in \eq{gen-funct} the generating functional
becomes
\begin{align}
  Z[J] = Z[0] \, {\rm exp}\left\{-\half\int d^4 x \, d^4 y \, J_\mu(x)
    \, {D}^{\mu\nu}(x, y) \, J_\nu(y) \right\} .
\label{prop-genf}
\end{align}
From (\ref{prop-genf}) we see that ${D}^{\mu\nu}(x,y)$ is indeed the
gluon propagator, as defined in (\ref{def-prop}), obtained in the
light-cone gauge with the sub-gauge condition (\ref{subgauge}).

We conclude that to find the gluon propagator we need to solve
\eq{greenf-eq} and verify that the solution leads to $a^\mu$
satisfying \eq{vanish-terms1}.

For any $x^- \ne \sigma$ the general solution of \eq{greenf-eq} is
\begin{align}
  D^{\mu\nu}(x,y)|_{x^-\ne \sigma} = \int \frac{d^4 k}{(2 \pi)^4} \,
  e^{-i k\cdot(x-y)} \, {-i\over k^2} \, \Big[g^{\mu\nu} -
  {k^\mu\tilden^\nu + k^\nu\tilden^\mu\over k^+}\Big],
\label{half-sol}
\end{align}
where the regularization of the $k^2 =0$ and $k^+ =0$ poles is not
specified on purpose, since the remaining uncertainty in this solution
is solely due to the freedom to regulate these poles in various ways.
Integrating \eq{greenf-eq} over $x^-$ in an infinitesimal interval
centered at $\sigma$ and assuming that $D^{\mu\nu}$ is continuous we
see that for $x^- = \sigma$ (and $y^- \ne \sigma$) the solution of
(\ref{greenf-eq}) has to satisfy the following condition
\begin{align}
  \partial^\perp_\mu \partial^\perp_\rho
  {D}^{\rho\nu}(x,y)|_{x^-=\sigma} = 0 ~ .
\label{cond-D}
\end{align}
(One also obtains continuity of $\partial_- D_{+\rho}$ at $x^- =
\sigma$.) The continuity of $ D^{\mu\nu}$ implies that its value at
$x^- = \sigma$ is fixed by \eq{half-sol}, such that we can write
\begin{align}
  D^{\mu\nu}(x,y) = \int \frac{d^4 k}{(2 \pi)^4} \, e^{-i k\cdot(x-y)}
  \, {-i\over k^2} \, \Big[g^{\mu\nu} - {k^\mu\tilden^\nu +
    k^\nu\tilden^\mu\over
    k^+}\Big] 
\label{D-sol}
\end{align}
for all $x^-$ with the only remaining freedom in this result being due
to unspecified regularization of the $k^2 =0$ and $k^+=0$ poles. In
fact one may still have different regularizations (or linear
combinations thereof) of the $k^2 =0$ and $k^+=0$ poles for $x^- >
\sigma$ and $x^- < \sigma$ in \eq{D-sol}. (For instance one may obtain
plane waves by replacing
\begin{align}
  \label{eq:repl}
  \frac{1}{k^2} \to \half \left[ \frac{1}{k^2 - i \epsilon} -
    \frac{1}{k^2 + i \epsilon} \right] = \pi \, i \, \delta (k^2)
\end{align}
in \eq{D-sol}.)  With the help of a direct calculation one can see
that no regularization of the $k^2 =0$ and $k^+=0$ poles in
\eq{half-sol} would lead to \eq{cond-D} for an arbitrary finite value
of $\sigma$ and for all $x^+, {\vec x}_\perp$. This leaves $\sigma =
\pm \infty$ as the only possibilities.

Let us first establish the Feynman prescription for the $k^2 =0$ pole
in \eq{D-sol}. Picking up the $x^+ = x_i^+$ and $x^+ = x_f^+$ surfaces
in \eq{vanish-terms1} and using $a^\mu$ from \eq{def-a} with the Green
function from \eq{D-sol} (with $k^2 \to k^2 + i \epsilon$) while
keeping in mind that $a^+=0$ in \eq{def-a} and $A^+ =0$ due to $\xi_1
\to 0$ limit in \eq{gen-funct} yields
\begin{subequations}
\begin{align}
  & \int dx^-\, d^2x_\perp \, A^\mu_ \perp(x^+_i)
  \Big(\sqrt{-(\partial^+)^2} +
  i \partial^+\Big)\,a^\perp_{\mu}(x^+_i)
  \nonumber\\
  & = \int \!\! d^4y \, dx^- \, d^2x_\perp \, A^\mu_\perp(x^+_i)\int
  \frac{d^4 k}{(2 \pi)^4} {2k^+\theta(k^+)\over k^2+i\epsilon}
  \Big(g^{\mu\nu}_\perp - {k^\mu_\perp\tilden^\nu\over k^+
  }\Big)\,e^{-i \, k^+(x^--y^-)-i \, k^-(x^+_i-y^+) +i \, {\vec
      k}_\perp \cdot ( {\vec x}_\perp - {\vec y}_\perp)} = 0
\label{Feyn-presc1}
\end{align}
and
\begin{align}
  & \int dx^-\, d^2x_\perp \, A_ \perp(x^+_f)^\mu
  \Big(\sqrt{-(\partial^+)^2} -
  i \partial^+\Big)\,a^\perp_{\mu}(x^+_f)
  \nonumber\\
  & = - \int\!\! d^4y \, dx^- \, d^2x_\perp A^\mu_\perp(x^+_f)\int
  \frac{d^4 k}{(2 \pi)^4} {2k^+\theta(-k^+)\over k^2+i\epsilon}
  \Big(g^{\mu\nu}_\perp - {k^\mu_\perp\tilden^\nu\over k^+}\Big)\,e^{-
    i \, k^+(x^--y^-)-ik^-(x^+_f-y^+) +i \, {\vec k}_\perp \cdot (
    {\vec x}_\perp - {\vec y}_\perp)} = 0.
\label{Feyn-presc2}
\end{align}
\end{subequations}
To prove the validity of Eqs. (\ref{Feyn-presc1}) and
(\ref{Feyn-presc2}), it is enough to observe that the direction of the
$k^-$ -contour closure is determined by the fact that $x^+_i - y^+< 0$
and $x^+_f - y^+>0$ for all $y^+$, since $x^+_i$ is the initial and
therefore the smallest $x^+$ value, while $x^+_f$ the final and
therefore the largest $x^+$ value in the 4-volume considered.
Eqs. (\ref{Feyn-presc1}) and (\ref{Feyn-presc2}) are zero independent
of the regularization prescription for the $k^+=0$ pole, and hence do
not allow us to fix this prescription. Note also that other
regularizations of the $k^2=0$ pole would not satisfy both
Eqs. (\ref{Feyn-presc1}) and (\ref{Feyn-presc2}).

We now write
\begin{align}
  D^{\mu\nu}(x,y) = \int \frac{d^4 k}{(2 \pi)^4} \, e^{-i k\cdot(x-y)}
  \, {-i\over k^2 + i \epsilon} \, \Big[g^{\mu\nu} - {k^\mu\tilden^\nu
    + k^\nu\tilden^\mu\over
    k^+}\Big] 
\label{D-sol1}
\end{align}
and directly face the need to regulate the $k^+ =0$ pole as the only
remaining ambiguity in the expression. Substituting \eq{D-sol1} into
\eq{cond-D} yields
\begin{align}
  \partial_\perp^\mu\int \frac{d^4 k}{(2 \pi)^4} \, {e^{-ik^+(\sigma -
      y^-) - i k^- (x^+ - y^+) + i \, {\vec k}_\perp \cdot ( {\vec
        x}_\perp - {\vec y}_\perp)} \over k^2+i\epsilon} \,
  \Big(k^\nu_\perp + {k^2_\perp \tilden^\nu \over [k^+]}\Big)\, = 0
\label{cond-D2}
\end{align}
where we have indicated with $[k^+]$ the prescription to be
determined. Once again we see that for finite $\sigma$ it is
impossible to satisfy \eq{cond-D2} and hence \eq{cond-D}.

Since $\sigma$ can not be finite, we consider $\sigma = + \infty$
first. In such case we need to close the $k^+$-integration contour in
the lower half-plane. Before doing the calculation, it is already
clear that our best chance of getting zero on the left-hand-side of
\eq{cond-D2} is to put $[k^+] = k^+ - i \epsilon$, such that the
light-cone pole would not contribute to the integral.

Using the following Fourier transform
\begin{align}
  &\int \frac{d^4 k}{(2 \pi)^4} \, e^{-ik\cdot (x-y)} {1\over
    k^2+i\epsilon}\Big(k^\nu_\perp + {k^2_\perp \tilden^\nu\over k^+ -
    i\epsilon}\Big)
  \nonumber\\
  & = {(x-y)^\nu_\perp\over 2\pi^2[(x-y)^2-i\epsilon]^2} + \tilden^\nu
  \left[{(x^- - y^-)\over \pi^2 [(x-y)^2-i\epsilon]^2} -
  i\delta^{(2)}({\vec x}_\perp - {\vec y}_\perp) \delta(x^+-y^+)
  \theta(y^- - x^-)\right]
\label{Fourier1}
\end{align}
we see that using $[k^+] = k^+ - i \epsilon$ satisfies \eq{cond-D2}
for $\sigma = + \infty$ since \eq{Fourier1} is zero for
$x^-=+\infty$. With this result we rewrite \eq{D-sol1} as
\begin{align}
  D^{\mu\nu}(x,y) = \int \frac{d^4 k}{(2 \pi)^4} \, e^{-i k\cdot(x-y)}
  \, {-i\over k^2 + i \epsilon} \, \left[g^{\mu\nu} -
  \frac{k^\mu\tilden^\nu}{k^+ - i \epsilon} - {k^\nu\tilden^\mu\over
    k^+}\right]. 
\label{D-sol2}
\end{align}
It may seem that there is still an unregulated pole at $k^+=0$ in the
last term of the square brackets in \eq{D-sol2}. However,
regularization of this last term can be fixed using the symmetry of
the gluon propagator, $D^{\mu\nu}(x,y) = D^{\nu\mu}(y,x)$. This yields
\begin{align}
  D^{\mu\nu}(x,y) = \int \frac{d^4 k}{(2 \pi)^4} \, e^{-i k\cdot(x-y)}
  \, {-i\over k^2 + i \epsilon} \, \left[g^{\mu\nu} -
  \frac{k^\mu\tilden^\nu}{k^+ - i \epsilon} - {k^\nu\tilden^\mu\over
    k^+ + i \epsilon}\right]. 
\label{D-sol3}
\end{align}

The derivation is similar for the case of $\sigma = - \infty$. We
employ
\begin{align}
  &\int \frac{d^4 k}{(2 \pi)^4} \, {1\over
    k^2+i\epsilon}\Big(k^\nu_\perp + {k^2_\perp \tilden^\nu\over k^+ +
    i\epsilon}\Big) \, e^{-ik\cdot (x-y)}
  \nonumber\\
  & = {(x-y)^\nu_\perp\over 2\pi^2[(x-y)^2-i\epsilon]^2} + \tilden^\nu
  \left[{(x^- - y^-)\over \pi^2 [(x-y)^2-i\epsilon]^2} + i
  \delta^{(2)}({\vec x}_\perp - {\vec y}_\perp)
  \delta(x^+-y^+)\theta(x^- - y^-)\right]
\label{Fourier2}
\end{align}
and observe that \eq{Fourier2} is zero for $x^- = -\infty$. Thus
\eq{cond-D2} is satisfied for $[k^+] = k^+ + i \epsilon$ and $\sigma =
- \infty$.

To summarize, we obtain the following two sub-gauge conditions and the
corresponding gluon propagators for $\sigma = \pm \infty$
\cite{Slavnov:1987yh,Kovchegov:1997pc,Belitsky:2002sm}:
\begin{itemize}
\item Light-cone gauge gluon propagator for the sub-gauge condition
  ${\vec \partial}_\perp \cdot {\vec A}_\perp (x^-=+\infty) = 0$
	\begin{align}
          D_1^{\mu\nu}(x,y) \equiv \int \frac{d^4 k}{(2 \pi)^4} \,
          e^{-ik\cdot(x-y)} \, {-i\over k^2+i\epsilon}\left[g^{\mu\nu}
            - {k^\mu \tilden^\nu \over k^+-i\epsilon} - {k^\nu
              \tilden^\mu \over k^++i\epsilon}\right] ;
	\label{propagator-1}
	\end{align}
\end{itemize}

\begin{itemize}
\item Light-cone gauge gluon propagator for the sub-gauge condition
  ${\vec \partial}_\perp \cdot {\vec A}_\perp (x^-=-\infty) = 0$
	\begin{align}
          D_2^{\mu\nu}(x,y) \equiv \int \frac{d^4 k}{(2 \pi)^4} \,
          e^{-ik\cdot(x-y)} \, {-i\over k^2+i\epsilon}
          \left[g^{\mu\nu} - {k^\mu \tilden^\nu \over k^++i\epsilon} -
            {k^\nu \tilden^\mu \over k^+-i\epsilon}\right].
	\label{propagator-2}
	\end{align}
\end{itemize}

As a consistency check, we now need to show that when using the
propagators (\ref{propagator-1}) or (\ref{propagator-2}),
Eq.~(\ref{vanish-terms1}) is satisfied along the $x^- = \pm \infty$
surfaces, along with the $x_\perp = \infty$ boundary. (We have checked
the $x^+ = x_i^+$ and $x^+ = x_f^+$ surfaces when deriving Feynman
regularization in Eqs.~\eqref{Feyn-presc1} and \eqref{Feyn-presc2}.)
\eq{vanish-terms1} is trivially satisfied at the $x_\perp = \infty$
boundary, since we assumed initially that the system is localized in
$x_\perp$ and all fields vanish when $x_\perp \to \infty$. We are left
only with the $x^- = \pm \infty$ surfaces to consider, for which
\eq{vanish-terms1} reduces to
\begin{align}
  - i \int d x^+ \, d^2x_\perp \, \Big[ A_\nu(\partial^- a^\nu) -
  A_\nu(\partial^\nu a^-)\Big] \Big|^{x^- = +\infty}_{x^- = -\infty}
  =0.
\label{vanish-terms2}
\end{align}
Let us demonstrate that \eq{vanish-terms2} is indeed valid for the
case of ${\vec \partial}_\perp \cdot {\vec A}_\perp(x^- = +\infty) =
0$ sub-gauge. (The argument for the ${\vec \partial}_\perp \cdot {\vec
  A}_\perp(x^- = -\infty) = 0$ sub-gauge is constructed by analogy.)
The $a^\mu$-shift is (cf. \eq{def-a})
\begin{align}
\label{a1}
a^\mu_1(x) = i \int d^4 y \, D_1^{\mu\nu}(x,y) \, J_\nu(y) ~.
\end{align}
We now plug \eq{a1} into \eq{vanish-terms2} and use \eq{propagator-1}
to integrate over $k^+$. Note that, just like in Eqs.~\eqref{Fourier1}
and \eqref{Fourier2}, picking up the $k^2=0$ pole of the
$k^+$-integral would give us a contribution which goes to zero as $x^-
\to \pm \infty$. (Those contributions are given by the first term on
the right-hand side of \eqref{Fourier1} and \eqref{Fourier2} and by
the first term in the square brackets of the right-hand side of
\eqref{Fourier1} and \eqref{Fourier2}.) Only picking the $k^+ =0$ pole
may give a term (akin to the last terms in the square brackets on the
right-hand side of \eqref{Fourier1} and \eqref{Fourier2}) which may
potentially violate \eq{vanish-terms2}. Therefore, we substitute
\eq{a1} into \eq{vanish-terms2} and use \eq{propagator-1} to integrate
over $k^+$ picking up the $k^+ =0$ poles only. Keeping in mind the
$A^+ =0$ gauge condition we write
\begin{align}
  & - i \int d x^+ \, d^2x_\perp \, \Big[ A_\nu(\partial^- a_1^\nu) -
  A_\nu(\partial^\nu a_1^-)\Big] \Big|^{x^- = +\infty}_{x^- = -\infty}
  \notag \\ & = \int d^4 y \, d x^+ \, d^2x_\perp \, J_\mu (y) \int
  \frac{d^4 k}{(2 \pi)^4} \, e^{-ik\cdot(x-y)} \, {-1 \over
    k^2+i\epsilon}\left[ k^- \, A^\mu (x) + k \cdot A (x) \,
    \frac{1}{k^+ + i \epsilon} \, \left( k^- \, \eta^\mu + k_\perp^\mu
    \right) \right]\Bigg|^{x^- = +\infty}_{x^- = -\infty} \notag \\ & =
  \int d^4 y \, d x^+ \, d^2x_\perp \, J_\mu (y) \int \frac{d^2
    k_\perp \, d k^-}{(2 \pi)^3} \, e^{-i k^- (x^+ - y^+) + i {\vec
      k}_\perp \cdot ({\vec x}_\perp - {\vec y}_\perp)} \, {i \over
    k_\perp^2} \, {\vec k}_\perp \cdot {\vec A}_\perp (x) \, \left(
    k^- \, \eta^\mu + k_\perp^\mu \right) \, \theta (x^- -
  y^-)\Big|^{x^- = +\infty}_{x^- = -\infty} \notag \\ & = \int d^4 y
  \, d x^+ \, d^2x_\perp \, J_\mu (y) \int \frac{d^2 k_\perp \, d
    k^-}{(2 \pi)^3} \, e^{-i k^- (x^+ - y^+) + i {\vec k}_\perp \cdot
    ({\vec x}_\perp - {\vec y}_\perp)} \, {i \over k_\perp^2} \, {\vec
    k}_\perp \cdot {\vec A}_\perp (x^- = + \infty) \, \left( k^- \,
    \eta^\mu + k_\perp^\mu \right) \notag \\ & = \int d^4 y \, d x^+
  \, d^2x_\perp \, J_\mu (y) \int \frac{d^2 k_\perp \, d k^-}{(2
    \pi)^3} \, e^{-i k^- (x^+ - y^+) + i {\vec k}_\perp \cdot ({\vec
      x}_\perp - {\vec y}_\perp)} \, {-1 \over k_\perp^2} \,
  {\vec \partial}_\perp \cdot {\vec A}_\perp (x^- = + \infty) \,
  \left( k^- \, \eta^\mu + k_\perp^\mu \right) =0,
\label{x-boundary}
\end{align}
where in the final steps we replaced ${\vec k}_\perp \to - i
{\vec \partial}_\perp$, integrated by parts, and employed the
${\vec \partial}_\perp \cdot {\vec A}_\perp(x^- = +\infty) = 0$
sub-gauge condition. The details of the calculation in \eq{x-boundary}
justifying neglecting the $k^2 =0$ pole in \eq{x-boundary} along with
the underlying assumptions are given in Appendix~\ref{A}. Note that
the contribution of the $k^2 =0$ pole is independent of the
regularization prescription for the $k^+ =0$ pole: hence the
conclusion of Appendix~\ref{A} is valid for all $k^+ =0$ pole
prescriptions.

Note that a 4-divergence sub-gauge condition, $\partial_\mu A^\mu (x^-
= +\infty) = 0$, would not have led to zero in \eq{x-boundary}, and
therefore does not correspond to propagator \eqref{propagator-1}. For
further reasons detailing why this is not a valid sub-gauge condition
of the light-cone gauge see Appendix~\ref{B}.

We have thus verified that $a^\mu$ from \eq{def-a} with either one of
the propagators \eqref{propagator-1} and \eqref{propagator-2}
satisfies \eq{vanish-terms1}, while the propagators
$D_1^{\mu\nu}(x,y)$ and $D_2^{\mu\nu}(x,y)$ solve \eq{greenf-eq} with
$\sigma = \pm \infty$ respectively. Therefore, \eq{prop-genf} is also
verified, with $D_1^{\mu\nu}(x,y)$ and $D_2^{\mu\nu}(x,y)$ being valid
light-cone gauge propagators satisfying corresponding sub-gauge
conditions.
 
It is also easy to explicitly check that propagators $D^{\mu\nu}_1$
and $D^{\mu\nu}_2$ themselves respect the sub-gauge conditions
\begin{align}
  &\partial_{\mu}^\perp D^{\mu\nu}_1(x,y)\Big|_{x^-=+\infty} = 0 ~,
  \nonumber \\
  &\partial_{\mu}^\perp D^{\mu\nu}_2(x,y)\Big|_{x^-=-\infty} = 0 ~.
\end{align}
 
Propagators (\ref{propagator-1}) and (\ref{propagator-2}) were already
obtained by different procedures in
\cite{Slavnov:1987yh,Kovchegov:1997pc,Belitsky:2002sm}. We observe
that in Ref. \cite{Belitsky:2002sm} the propagators
(\ref{propagator-1}) and (\ref{propagator-2}) were obtained by
imposing an additional sub-gauge condition, $A^-(x^-=\pm\infty) =0$,
while in the above procedure we showed that it is sufficient to assume
that $\lim\limits_{x^- \to \infty} \left[ A^-(x^-)/x^- \right] =0$
(see Appendix~\ref{A}).
 

\section{PV Sub-Gauge}
\label{sec:PV}

In this section we will determine the sub-gauge condition that
reproduces Principal Value (PV) prescription \eqref{propagator-PV0}
for the $k^+$ pole in light-cone propagator.  To this end, we will
adopt the same procedure we used to arrive at propagators
(\ref{propagator-1}) and (\ref{propagator-2}) with sub-gauge
conditions ${\vec \partial}_\perp\cdot {\vec A}_\perp (x^-=+\infty)
=0$ and ${\vec \partial}_\perp \cdot {\vec A}_\perp (x^-=-\infty) =0$
respectively, but in reverse order.

In the previous section we have assumed a sub-gauge condition
\eqref{subgauge}, performed a shift of the field $A^\mu \to A^\mu +
a^\mu$ in the generating functional, and made sure that the
$a^\mu$-dependent surface terms vanish (that is, \eq{vanish-terms1} is
satisfied) for the generating functional to reduce to the form given
in (\ref{prop-genf}).

As we do not know \textit{a priori} the sub-gauge condition that
reproduces the light-cone propagator with $k^+=0$ pole regulated by PV
prescription, we consider from the start the propagator with the PV
prescription and deduce the needed sub-gauge condition in order to put
the generating functional in the form (\ref{prop-genf}). In practical
terms, we have to show that Eq. (15) is satisfied if we regulate the
$k^+=0$ pole of the light-cone propagator with the ${\rm PV}$
prescription.

The gauge field propagator in the $A^+=0$ light-cone gauge with the
${\rm PV}$-prescription is
\begin{align}
  D_{PV}^{\mu\nu}(x, y) \equiv \int \frac{d^4 k}{(2 \pi)^4} \,
  e^{-ik\cdot(x-y)} \, {-i\over k^2+i\epsilon} \left[ g^{\mu\nu} -
    \Big(k^\mu \tilden^\nu + k^\nu \tilden^\mu\Big){\rm
      PV}\left\{{1\over k^+}\right\} \right]
\label{propagator-PV}
\end{align}
where
\begin{align} {\rm PV}\left\{{1\over k^+}\right\} \equiv \half
  \left({1\over k^+-i\epsilon} + {1 \over k^++i\epsilon}\right).
\end{align}
Knowing the propagator means we know the shift field $a^\mu$
(cf. \eq{def-a}),
\begin{align}
\label{PV-a}
a^\mu_{PV} = i \int d^4 y \, D^{\mu\nu}_{PV}(x,y) \, J_\nu(y).
\end{align}
Let us plug the shift field \eqref{PV-a} into
Eq.~(\ref{vanish-terms1}) obtaining
\begin{align}
  \int d x^- \, d^2x_\perp \, \Big( A_i^\mu \sqrt{-(\partial^+)^2} \,
  a^{PV}_{i\,\mu} + A_f^\mu \sqrt{-(\partial^+)^2} \,
  a^{PV}_{f\,\mu}\Big) - i \int d\sigma_\mu \Big[ A_\nu(\partial^\mu
  a_{PV}^\nu) - A_\nu(\partial^\nu a_{PV}^\mu)\Big] =0
\label{vanish-terms3}
\end{align}
and require that the latter is satisfied everywhere along the boundary
of the four-dimensional space-time volume. \eq{vanish-terms3} is
satisfied at the $x^+ = x^+_i$ and $x^+ = x^+_f$ boundaries
irrespective of the regularization of the $k^+ =0$ pole, as follows
from Eqs.~\eqref{Feyn-presc1} and \eqref{Feyn-presc2}. The boundary at
$x_\perp \to \infty$ is also automatically satisfied, since we assumed
from the start that all fields vanish as $x_\perp \to \infty$.  We are
only left with the boundary at $x^- = \pm \infty$. By analogy to
\eq{x-boundary} we evaluate the contributions of the $x^- = \pm
\infty$ boundaries by neglecting the residues of $k^2 =0$ pole in the
propagator which vanish at those boundaries (see Appendix~\ref{A} and
Eqs.~\eqref{Fourier1} and \eqref{Fourier2}):
\begin{align}
  & 0 = - i \int d x^+ \, d^2x_\perp \, \Big[ A_\nu(\partial^-
  a_{PV}^\nu) - A_\nu(\partial^\nu a_{PV}^-)\Big] \Big|^{x^- =
    +\infty}_{x^- = -\infty} \notag \\ & = \int d^4 y \, d x^+ \,
  d^2x_\perp \, J_\mu (y) \int \frac{d^4 k}{(2 \pi)^4} \,
  e^{-ik\cdot(x-y)} \, {-1 \over k^2+i\epsilon}\left[ k^- \, A^\mu (x)
    + k \cdot A (x) \, {\rm PV}\left\{\frac{1}{k^+ + i
        \epsilon}\right\} \, \left( k^- \, \eta^\mu + k_\perp^\mu
    \right) \right]\Bigg|^{x^- = +\infty}_{x^- = -\infty} \notag \\ &
  = \int d^4 y \, d x^+ \, d^2x_\perp \, J_\mu (y) \int \frac{d^2
    k_\perp \, d k^-}{(2 \pi)^3} \, e^{-i k^- (x^+ - y^+) + i {\vec
      k}_\perp \cdot ({\vec x}_\perp - {\vec y}_\perp)} \, {i \over
    k_\perp^2} \, {\vec k}_\perp \cdot {\vec A}_\perp (x) \, \left(
    k^- \, \eta^\mu + k_\perp^\mu \right) \, \half \, {\rm Sign} (x^-
  - y^-)\Big|^{x^- = +\infty}_{x^- = -\infty} \notag \\ & = \int d^4 y
  \, d x^+ \, d^2x_\perp \, J_\mu (y) \int \frac{d^2 k_\perp \, d
    k^-}{2 (2 \pi)^3} \, e^{-i k^- (x^+ - y^+) + i {\vec k}_\perp
    \cdot ({\vec x}_\perp - {\vec y}_\perp)} \, {i \over k_\perp^2} \,
  \left[ {\vec k}_\perp \cdot {\vec A}_\perp (x^- = + \infty) + {\vec
      k}_\perp \cdot {\vec A}_\perp (x^- = - \infty) \right] \notag \\
  & \times \, \left( k^- \, \eta^\mu + k_\perp^\mu \right) = \int d^4
  y \, d x^+ \, d^2x_\perp \, J_\mu (y) \int \frac{d^2 k_\perp \, d
    k^-}{2 (2 \pi)^3} \, e^{-i k^- (x^+ - y^+) + i {\vec k}_\perp
    \cdot ({\vec x}_\perp - {\vec y}_\perp)} \, {-1 \over k_\perp^2}
  \, \left( k^- \, \eta^\mu + k_\perp^\mu \right) \notag \\ & \times
  \left[ {\vec \pd}_\perp \cdot {\vec A}_\perp (x^- = + \infty) +
    {\vec \pd}_\perp \cdot {\vec A}_\perp (x^- = - \infty) \right].
\label{x-boundaryPV}
\end{align}

We see that for the boundary condition in \eq{x-boundaryPV} to be
satisfied, i.e. for the boundary term to vanish, one has to have the
following sub-gauge condition:
\begin{align}
  {\vec \partial}_\perp\cdot {\vec A}_\perp(x^-=+\infty) +
  {\vec \partial}_\perp\cdot {\vec A}_\perp(x^-=-\infty) =0.
\label{PV-subgauge}
\end{align}
We have thus arrived at the sub-gauge condition which leads to the
$k^+$ pole in the gluon propagator regulated with the ${\rm PV}$
prescription. We can check the validity of the PV-sub-gauge condition
(\ref{PV-subgauge}) explicitly by using the PV-propagator:
\begin{align}
  \partial_{\mu}^\perp D^{\mu\nu}_{PV}(x,y)\Big|_{x^-=+\infty}
  +\partial_{\mu}^\perp D^{\mu\nu}_{PV}(x,y)\Big|_{x^-=-\infty} = 0.
\end{align}

In Section~\ref{Class} we will show that the PV sub-gauge condition
(\ref{PV-subgauge}) is consistent with reproducing the classical gluon
field generated by two ultrarelativistic quarks propagating along two
parallel light-cones, whereas a stronger condition
\begin{align} 
\label{PVwrong}
  {\vec A}_\perp(x^-=+\infty) + {\vec A}_\perp(x^-=-\infty) =0,
\end{align}
while still satisfying \eq{x-boundaryPV} does not allow one to
construct the classical field of the color charges at the non-Abelian
level. Therefore, it is \eq{PV-subgauge} which appears to be the
correct sub-gauge condition in the PV case.


\section{Mandelstam--Leibbrandt prescription}
\label{sec:ML}

In this section we will try to obtain the sub-gauge condition that is
consistent with the light-cone gauge propagator \eqref{ML-propag0}
with $k^+=0$ pole regulated by Mandelstam-Leibbrandt (ML) prescription
\cite{Mandelstam:1982cb,Leibbrandt:1983pj}. To this end, we will adopt
the same procedure we used for the PV sub-gauge in the previous
Section, \textit{i.e}, we will use the ML propagator
\eqref{ML-propag0} to construct the shift field $a^\mu$ from
\eqref{def-a}, and use the latter in Eq.~(\ref{vanish-terms1}) to try
to deduce the sub-gauge condition that has to be satisfied.

The light-cone propagator with Mandelstam-Leibbrandt prescription
\cite{Mandelstam:1982cb,Leibbrandt:1983pj} is
\begin{align}
  D^{\mu\nu}_{ML}(x,y) = \int \!\! \frac{d^4 k}{(2 \pi)^4} \, e^{-i
    k\cdot(x-y)} \, {-i\over k^2+i\epsilon} \left[ g^{\mu\nu} -
    {k^\mu\tilden^\nu + k^\nu\tilden^\mu \over k^+ + i\epsilon k^-}
  \right] ~.
\label{ML-propag}
\end{align}
The corresponding shift field is
\begin{align}
\label{aML}
a^\mu_{ML} = i \, \int d^4 y \, D^{\mu\nu}_{ML}(x, y) \, J_\nu(y).
\end{align}
Substituting $a^\mu_{ML}$ into \eq{vanish-terms1} yields the following
boundary condition for $a^\mu_{ML}$ to satisfy:
\begin{align}
  \int d x^- \, d^2x_\perp \, \Big( A_i^\mu \sqrt{-(\partial^+)^2} \,
  a^{ML}_{i\,\mu} + A_f^\mu \sqrt{-(\partial^+)^2} \,
  a^{ML}_{f\,\mu}\Big) - i \int d\sigma_\mu \Big[ A_\nu(\partial^\mu
  a_{ML}^\nu) - A_\nu(\partial^\nu a_{ML}^\mu)\Big] =0.
\label{vanish-terms4}
\end{align}
Again only the $x^- = \pm \infty$ boundaries need to be considered,
since the other boundary conditions are automatically satisfied by the
field from \eq{aML}. Discarding the contributions of the $k^2 =0$ pole
we get
\begin{align}
  & 0 = - i \int d x^+ \, d^2x_\perp \, \Big[ A_\nu(\partial^-
  a_{ML}^\nu) - A_\nu(\partial^\nu a_{ML}^-)\Big] \Big|^{x^- =
    +\infty}_{x^- = -\infty} \notag \\ & = \int d^4 y \, d x^+ \,
  d^2x_\perp \, J_\mu (y) \int \frac{d^4 k}{(2 \pi)^4} \,
  e^{-ik\cdot(x-y)} \, {-1 \over k^2+i\epsilon}\left[ k^- \, A^\mu (x)
    + k \cdot A (x) \, \frac{1}{k^+ + i \epsilon k^-} \, \left( k^- \,
      \eta^\mu + k_\perp^\mu \right) \right]\Bigg|^{x^- =
    +\infty}_{x^- = -\infty} \notag \\ & = \int d^4 y \, d x^+ \,
  d^2x_\perp \, J_\mu (y) \int \frac{d^2 k_\perp \, d k^-}{(2 \pi)^3}
  \, e^{-i k^- (x^+ - y^+) + i {\vec k}_\perp \cdot ({\vec x}_\perp -
    {\vec y}_\perp)} \, {i \over k_\perp^2} \, {\vec k}_\perp \cdot
  {\vec A}_\perp (x) \, \left( k^- \, \eta^\mu + k_\perp^\mu \right)
  \notag \\ & \times \, \half \, \left[ \theta (x^- - y^-) \, \theta
    (k^-) - \theta (y^- - x^-) \, \theta (-k^-) \right] \Big|^{x^- =
    +\infty}_{x^- = -\infty} \notag \\ & = \int d^4 y \, d x^+ \,
  d^2x_\perp \, J_\mu (y) \int \frac{d^2 k_\perp \, d k^-}{2 (2
    \pi)^3} \, e^{-i k^- (x^+ - y^+) + i {\vec k}_\perp \cdot ({\vec
      x}_\perp - {\vec y}_\perp)} \, {i \over k_\perp^2} \, \left( k^-
    \, \eta^\mu + k_\perp^\mu \right) \notag \\ & \times \, \left[
    \theta (k^-) \, {\vec k}_\perp \cdot {\vec A}_\perp (x^- = +
    \infty) + \theta (- k^-) \, {\vec k}_\perp \cdot {\vec A}_\perp
    (x^- = - \infty) \right] \notag \\ & = \int d^4 y \, d x^+ \,
  d^2x_\perp \, J_\mu (y) \int \frac{d^2 k_\perp \, d k^-}{2 (2
    \pi)^3} \, e^{-i k^- (x^+ - y^+) + i {\vec k}_\perp \cdot ({\vec
      x}_\perp - {\vec y}_\perp)} \, {-1 \over k_\perp^2} \, \left(
    k^- \, \eta^\mu + k_\perp^\mu \right) \notag \\ & \times \left[
    \theta (k^-) \, {\vec \pd}_\perp \cdot {\vec A}_\perp (x^- = +
    \infty) + \theta (- k^-) \, {\vec \pd}_\perp \cdot {\vec A}_\perp
    (x^- = - \infty) \right].
\label{x-boundaryML}
\end{align}
It appears that to satisfy the boundary condition we need to require
that the expression in the last square brackets in \eq{x-boundaryML}
is zero. However, the expression in the square brackets depends on
$k^-$: equating it to zero would result in a sub-gauge condition which
would depend on the arbitrary momentum $k^-$, mixing up coordinate and
momentum spaces. Such condition can only be satisfied if each term in
the last square brackets of \eq{x-boundaryML} is zero separately.

The situation does not change if we integrate over $k^-$ in
\eq{x-boundaryML} obtaining
\begin{align}
  & 0 = \int d^4 y \, d x^+ \, d^2x_\perp \, J_\mu (y) \int \frac{d^2
    k_\perp \, d k^-}{2 (2 \pi)^3} \, e^{i {\vec k}_\perp \cdot ({\vec
      x}_\perp - {\vec y}_\perp)} \, {-1 \over k_\perp^2} \, \left[ -
    \left( \frac{\eta^\mu}{(x^+ - y^+ - i \epsilon)^2} + \frac{i
        k_\perp^\mu}{x^+ - y^+ - i \epsilon} \right) \, {\vec
      \pd}_\perp \cdot {\vec A}_\perp (x^- = + \infty) \right. \notag
  \\ & \left. + \left( \frac{\eta^\mu}{(x^+ - y^+ + i \epsilon)^2} +
      \frac{i k_\perp^\mu}{x^+ - y^+ + i \epsilon} \right) \, {\vec
      \pd}_\perp \cdot {\vec A}_\perp (x^- = - \infty) \right].
    \label{x-boundaryML2}
\end{align}
The two terms in the square brackets of \eq{x-boundaryML2} are
multiplied by two different functions of an arbitrary variable
$y^+$. Again the only way for these square brackets to be equal to
zero is to require that
\begin{align}
  & {\vec \partial}_\perp \cdot {\vec A}_\perp (x^-=+\infty)=0 \ \ \
  \mbox{and}
  \nonumber\\
  & {\vec \partial}_\perp \cdot {\vec A}_\perp (x^-=-\infty)=0
\label{ML-subg}
\end{align}
at the same time. However, the sub-gauge conditions (\ref{ML-subg})
are not satisfied by the ML-propagator (\ref{ML-propag}). Indeed, we
have
\begin{subequations}\label{propML}
\begin{align}
  & \partial_{\mu}^\perp D_{ML}^{\mu\nu}(x, y)|_{x^-=+\infty} = -
  \frac{1}{2 \pi} \, \eta^\nu \, \delta^{(2)} \left( {\vec x}_\perp -
    {\vec y}_\perp \right) \, \frac{1}{x^+-y^+ - i\epsilon} \ne 0
  \\
  & \partial_{\mu}^\perp D_{ML}^{\mu\nu}(x, y)|_{x^-=-\infty} = -
  \frac{1}{2 \pi} \, \eta^\nu \, \delta^{(2)} \left( {\vec x}_\perp -
    {\vec y}_\perp \right) \, \frac{1}{x^+-y^+ + i\epsilon} \ne 0 ~.
\end{align}
\end{subequations}
In addition, the conditions (\ref{ML-subg}) can not even be satisfied
by the classical gluon field of a single relativistic charge, as will
become apparent in Section~\ref{Class}.

For $x^+ \neq y^+$ \eq{x-boundaryML2} can be satisfied by requiring
that
\begin{align}
\label{MLconstr}
{\vec \partial}_\perp \cdot {\vec A}_\perp (x^-=+\infty) =
{\vec \partial}_\perp \cdot {\vec A}_\perp (x^-=-\infty).
\end{align}
However, there is no reason here to require $x^+ \neq y^+$, since both
variables are integrated over in \eqref{x-boundaryML2}. In addition,
\eq{MLconstr} is not satisfied even by the field of a single
ultrarelativistic charge in electrodynamics. Finally, even the ML
propagators do not satisfy \eqref{MLconstr}, as can be seen from
\eqref{propML}.

We conclude that the procedure with which we successfully determined
the sub-gauge condition for PV-prescription is either not the right
procedure to obtain the sub-gauge condition for the light-cone gluon
propagator with ML-prescription or that the ML light-cone propagator
is not compatible with the functional integral formalism.

It is interesting to observe that, in \cite{Slavnov:1987yh} the
ML-light-cone propagator was obtained within the functional integral
formalism using complex valued fields (for a real-field gauge theory).


\section{Classical Yang-Mills Field}
\label{Class}

In this Section we illustrate the PV sub-gauge condition
\eqref{PV-subgauge} and the corresponding propagator
\eqref{propagator-PV} by an example of a classical gluon field of two
color charges on parallel light cones calculated in the $A^+ =0$
light-cone gauge. This types of problems arise in the
McLerran--Venugopalan (MV) model
\cite{McLerran:1993ni,McLerran:1993ka,McLerran:1994vd,Kovchegov:1996ty,Kovchegov:1997pc,Jalilian-Marian:1997xn}
of a large nucleus, where the classical gluon field dominates over
quantum corrections due to small coupling and large atomic number of
the nucleus (see \cite{KovchegovLevin} for a detailed introduction to
the subject). Classical gluon field of a single ultrarelativistic
nucleus in the $\theta$-function sub-gauges of the $A^+ =0$ light-cone
gauge was constructed in the MV model framework by solving Yang-Mills
(YM) equations in \cite{Kovchegov:1996ty,Jalilian-Marian:1997xn} and
by summation of the corresponding tree-level diagrams in
\cite{Kovchegov:1997pc}. Below we will repeat both types of
calculations for the PV sub-gauge of the $A^+ =0$ light-cone gauge for
a system of two color charges, which could be two valence quarks from
two nucleons in a large nucleus.\footnote{Note that since above we
  have failed to find the sub-gauge condition corresponding to the ML
  prescription, we can not solve classical YM equations in the ML
  case, since we do not know which condition to impose on the field. A
  diagrammatic calculation with the ML gluon propagator should lead to
  the field equivalent to one of the $\theta$-function sub-gauges
  since $k^- < 0$ for all virtual gluon lines in this case.} The
calculations in this Section closely follows what was done in
\cite{Kovchegov:1996ty,Kovchegov:1997pc}, but in a different sub-gauge
of the light-cone gauge.

Consider two ultrarelativistic quarks on two parallel light-cones. In
covariant (Feynman) $\pd_\mu A^\mu =0$ gauge their classical gluon
field is known exactly \cite{Kovchegov:1996ty,Jalilian-Marian:1997xn}
and is
\begin{align}
  \label{Acov}
  A_{cov}^{a \, +} (x^-, {\vec x}_\perp) = \frac{g}{2 \pi} \, (t^a)_1
  \, \delta (x^- - b_{1}^- ) \, \ln \left(|{\vec x}_\perp - {\vec
      b}_{1\perp}| \, \Lambda \right) + \frac{g}{2 \pi} \, (t^a)_2 \,
  \delta (x^- - b_{2}^- ) \, \ln \left(|{\vec x}_\perp - {\vec
      b}_{2\perp}| \, \Lambda \right),
\end{align}
where $(b_{1}^- , {\vec b}_{1\perp})$ and $(b_{2}^- , {\vec
  b}_{2\perp})$ determine the quarks' light-cone trajectories, $g$ is
the coupling constant, and $(t^a)_i$ are fundamental SU($N_c$)
generators in the color space of quark $i$.

We need to find the gauge transformation from Feynman to the
light-cone gauge. It is given by
\begin{align}
  \label{eq:gauge_tr_LC}
  A_\mu^{LC} = S \, A_\mu^{cov} \, S^{-1} - \frac{i}{g} \,
  (\partial_\mu S) \, S^{-1}. 
\end{align}
Requiring that the new gauge is the light-cone gauge, $A^+_{LC} =0$,
yields the following differential equation:
\begin{align}
  \label{eq:LCgaugeS}
  \partial^+ S = - i \, g \, S \, A^+_{cov}.
\end{align}
As discussed in the Introduction, \eq{eq:LCgaugeS} does not specify
$S$, and hence the gauge, uniquely. In the PV sub-gauge it needs to be
augmented by the boundary condition \eqref{PV-subgauge}.

While \eq{Acov} is the exact solution of the Yang-Mills equations for
two ultrarelativistic charges, we will try to construct $S$ by solving
\eq{eq:LCgaugeS} order-by-order in $g^2$, making sure the condition
\eqref{PV-subgauge} is satisfied by the light-cone gauge gluon field
at each order.


\subsection{Abelian Case}

$S$ is a unitary matrix. At the lowest non-trivial order we write 
\begin{align}
  \label{eq:S_LO}
  S = 1 + i \, \alpha (x^-, {\vec x}_\perp) + \ldots , 
\end{align}
where $\alpha (x)$ is an order-$g^2$ correction and ellipsis represent
higher-order corrections in $g$. Since $S$ is unitary, $\alpha (x)$ is
a hermitean matrix. Plugging \eq{eq:S_LO} into \eq{eq:LCgaugeS} we get
\begin{align}
  \label{eq:diff1}
  \partial^+ \alpha = - g \, A^+_{cov}. 
\end{align}
Solving this equation with $A^+_{cov}$ given by \eq{Acov} we obtain
\begin{align}
  \label{eq:a_LO}
  \alpha (x^-, {\vec x}_\perp) = - \frac{g^2}{2 \pi} \, t^a (t^a)_1 \,
  \frac{1}{2} \, \mbox{Sign} (x^- - b_{1}^- ) \, \ln \left(|{\vec
      x}_\perp - {\vec b}_{1\perp}| \, \Lambda \right) - \frac{g^2}{2
    \pi} \, t^a (t^a)_2 \, \frac{1}{2} \, \mbox{Sign} (x^- - b_{2}^- )
  \, \ln \left(|{\vec x}_\perp - {\vec b}_{2\perp}| \, \Lambda \right)
  \notag \\ + C_1 ({\vec x}_\perp, b_1, b_2),
\end{align}
where $C_1$ is the integration constant (which may be a function of
all the other variables in the problem). To find $C_1$ we need to
satisfy the boundary condition \eqref{PV-subgauge}. The transverse
components of the gluon field in the LC gauge are given by (note that
$\partial^i_\perp = - \nabla_\perp^i$)
\begin{align}
  \label{eq:A_LC_LO}
  {\vec A}_\perp^{LC} (x^-, {\vec x}_\perp) = \frac{i}{g} \, ({\vec
    \nabla}_\perp S) \, S^{-1} = - \frac{1}{g} \,{\vec \nabla}_\perp
  \alpha (x^-, {\vec x}_\perp) + \ldots .
\end{align}

Using \eq{eq:a_LO} in \eq{eq:A_LC_LO} yields
\begin{align}
  \label{eq:A_LC_LO1}
  {\vec A}_\perp^{LC} (x^-, {\vec x}_\perp) = \frac{g}{2 \pi} \, t^a
  (t^a)_1 \, \frac{1}{2} \, \mbox{Sign} (x^- - b_{1}^- ) \,
  \frac{{\vec x}_\perp - {\vec b}_{1\perp}}{|{\vec x}_\perp - {\vec
      b}_{1\perp}|^2} + \frac{g}{2 \pi} \, t^a (t^a)_2 \, \frac{1}{2}
  \, \mbox{Sign} (x^- - b_{2}^- ) \, \frac{{\vec x}_\perp - {\vec
      b}_{2\perp}}{|{\vec x}_\perp - {\vec b}_{2\perp}|^2} \notag \\ -
  \frac{1}{g} \, {\vec \nabla}_\perp C_1 ({\vec x}_\perp, b_1, b_2) +
  \ldots .
\end{align}
Clearly the gluon field from \eq{eq:A_LC_LO} satisfies the condition
\eqref{PV-subgauge} iff
\begin{align}
  \label{eq:C1}
  {\vec \nabla}_\perp C_1 ({\vec x}_\perp, b_1, b_2) = 0,
\end{align}
which means that $C_1  = C_1 (b_1, b_2)$,
\begin{align}
  \label{eq:A_LC_LO2}
  {\vec A}_\perp^{LC} (x^-, {\vec x}_\perp) = \frac{g}{2 \pi} \, t^a
  (t^a)_1 \, \frac{1}{2} \, \mbox{Sign} (x^- - b_{1}^- ) \,
  \frac{{\vec x}_\perp - {\vec b}_{1\perp}}{|{\vec x}_\perp - {\vec
      b}_{1\perp}|^2} + \frac{g}{2 \pi} \, t^a (t^a)_2 \, \frac{1}{2}
  \, \mbox{Sign} (x^- - b_{2}^- ) \, \frac{{\vec x}_\perp - {\vec
      b}_{2\perp}}{|{\vec x}_\perp - {\vec b}_{2\perp}|^2} + {\cal O} (g^3)
\end{align}
and 
\begin{align}
  \label{eq:a_LO22}
  \alpha (x^-, {\vec x}_\perp) = - \frac{g^2}{2 \pi} \, t^a (t^a)_1 \,
  \frac{1}{2} \, \mbox{Sign} (x^- - b_{1}^- ) \, \ln \left(|{\vec
      x}_\perp - {\vec b}_{1\perp}| \, \Lambda \right) - \frac{g^2}{2
    \pi} \, t^a (t^a)_2 \, \frac{1}{2} \, \mbox{Sign} (x^- - b_{2}^- )
  \, \ln \left(|{\vec x}_\perp - {\vec b}_{2\perp}| \, \Lambda \right)
  \notag \\ + C_1 (b_1, b_2).
\end{align}
Furthermore, since the field is Abelian at this order, the function
$C_1$ is additive, $C_1 (b_1, b_2) = {\tilde C} (b_1) + {\tilde C}
(b_2)$. Applying translational invariance gives ${\tilde C}
(b)$=const, while this constant we will put to zero. (The appearance
of the function $C_1$ is related to the fact that even our sub-gauge
conditions do not fix the field uniquely: an Abelian gauge
transformation \eqref{eq:gauge_tr} with $\nabla_\perp^2 \Lambda (x^+,
{\vec x}_\perp) =0$ preserves both the light-cone gauge and the
sub-gauge condition \eqref{PV-subgauge}.)

Without $C_1$ we write
\begin{align}
  \label{eq:a_LO2}
  \alpha (x^-, {\vec x}_\perp) = - \frac{g^2}{2 \pi} \, t^a (t^a)_1 \,
  \frac{1}{2} \, \mbox{Sign} (x^- - b_{1}^- ) \, \ln \left(|{\vec
      x}_\perp - {\vec b}_{1\perp}| \, \Lambda \right) - \frac{g^2}{2
    \pi} \, t^a (t^a)_2 \, \frac{1}{2} \, \mbox{Sign} (x^- - b_{2}^- )
  \, \ln \left(|{\vec x}_\perp - {\vec b}_{2\perp}| \, \Lambda
  \right).
\end{align}


\subsection{Non-Abelian Corrections}

Let us find the next correction to $S$. Remembering that $S$ is
unitary we write
\begin{align}
  \label{eq:S_NLO}
  S = 1 + i \, \alpha - \frac{\alpha^2}{2} + i \, \alpha' + \ldots ,
\end{align}
where $\alpha'(x)$ is the order-$g^4$ correction, which again is a
hermitean matrix. Plugging \eqref{eq:S_NLO} into \eq{eq:LCgaugeS} and
employing \eq{eq:diff1} yields
\begin{align}
  \label{eq:aprime}
  \partial^+ \alpha' = \frac{i}{2} \, \left[ \alpha, \partial^+ \alpha
  \right].
\end{align}
Using \eq{eq:a_LO2} in \eqref{eq:aprime} we write
\begin{align}
  \label{eq:aprime_long}
  \partial^+ \alpha' = & \, \frac{i}{2} \left( \frac{g^2}{2
      \pi}\right)^2 \, [t^a (t^a)_1 , t^b (t^b)_2 ] \, \ln
  \left(|{\vec x}_\perp - {\vec b}_{1\perp}| \, \Lambda \right) \, \ln
  \left(|{\vec x}_\perp - {\vec b}_{2\perp}| \, \Lambda \right) \,
  \frac{1}{2} \, \mbox{Sign} (b_2^- - b_{1}^- ) \, \left[ \delta (x^-
    - b_{2}^- ) + \delta (x^- - b_{1}^- ) \right].
\end{align}
The solution of \eq{eq:aprime_long} is
\begin{align}
  \label{eq:aprime_sol}
  \alpha' = & \, \frac{i}{8} \left( \frac{g^2}{2 \pi}\right)^2 \, [t^a
  (t^a)_1 , t^b (t^b)_2 ] \, \ln \left(|{\vec x}_\perp - {\vec
      b}_{1\perp}| \, \Lambda \right) \, \ln \left(|{\vec x}_\perp -
    {\vec b}_{2\perp}| \, \Lambda \right) \, \mbox{Sign} (b_2^- -
  b_{1}^- ) \, \left[ \mbox{Sign} (x^- - b_{2}^- ) + \mbox{Sign} (x^-
    - b_{1}^- ) \right] \notag \\ & + C_2 ({\vec x}_\perp, b_1, b_2)
\end{align}
with $C_2$ the integration constant. 

To impose the sub-gauge condition \eqref{PV-subgauge} we need to find
the transverse components of the gluon field in the light-cone
gauge. We write
\begin{align}
  \label{eq:A_LC_NLO}
  {\vec A}_\perp^{LC} (x^-, {\vec x}_\perp) = \frac{i}{g} \, ({\vec
    \nabla}_\perp S) \, S^{-1} = - \frac{1}{g} \, {\vec \nabla}_\perp
  \alpha - \frac{1}{g} \, {\vec \nabla}_\perp \alpha' - \frac{i}{2 g}
  [\alpha, {\vec \nabla}_\perp \alpha] + \ldots .
\end{align}
Substituting Eqs.~\eqref{eq:a_LO2} and \eqref{eq:aprime_sol} into
\eq{eq:A_LC_NLO} gives
\begin{align}
  \label{eq:A_LC_NLOlong}
  {\vec A}_\perp^{LC} (x^-, {\vec x}_\perp) & = \frac{g}{4 \pi} \, t^a
  (t^a)_1 \, \mbox{Sign} (x^- - b_{1}^- ) \, \frac{{\vec x}_\perp -
    {\vec b}_{1\perp}}{|{\vec x}_\perp - {\vec b}_{1\perp}|^2} +
  \frac{g}{4 \pi} \, t^a (t^a)_2 \, \mbox{Sign} (x^- - b_{2}^- ) \,
  \frac{{\vec x}_\perp - {\vec b}_{2\perp}}{|{\vec x}_\perp - {\vec
      b}_{2\perp}|^2} \notag \\ & - \frac{i}{8} \frac{g^3}{(2 \pi)^2}
  \, [t^a (t^a)_1 , t^b (t^b)_2 ] \, \mbox{Sign} (b_2^- - b_{1}^- ) \,
  \left[ \mbox{Sign} (x^- - b_{2}^- ) + \mbox{Sign} (x^- - b_{1}^- )
  \right] \notag \\ & \times \left[ \frac{{\vec x}_\perp - {\vec
        b}_{1\perp}}{|{\vec x}_\perp - {\vec b}_{1\perp}|^2} \, \ln
    \left(|{\vec x}_\perp - {\vec b}_{2\perp}| \, \Lambda \right) +
    \frac{{\vec x}_\perp - {\vec b}_{2\perp}}{|{\vec x}_\perp - {\vec
        b}_{2\perp}|^2} \, \ln \left(|{\vec x}_\perp - {\vec
        b}_{1\perp}| \, \Lambda \right) \right] \notag \\ & -
  \frac{1}{g} \, {\vec \nabla}_\perp C_2 ({\vec x}_\perp, b_1, b_2) -
  \frac{i}{8} \frac{g^3}{(2 \pi)^2} \, [t^a (t^a)_1 , t^b (t^b)_2 ] \,
  \mbox{Sign} (x^- - b_{1}^- ) \, \mbox{Sign} (x^- - b_{2}^- ) \notag
  \\ & \times \left[ \frac{{\vec x}_\perp - {\vec b}_{2\perp}}{|{\vec
        x}_\perp - {\vec b}_{2\perp}|^2} \, \ln \left(|{\vec x}_\perp
      - {\vec b}_{1\perp}| \, \Lambda \right) - \frac{{\vec x}_\perp -
      {\vec b}_{1\perp}}{|{\vec x}_\perp - {\vec b}_{1\perp}|^2} \,
    \ln \left(|{\vec x}_\perp - {\vec b}_{2\perp}| \, \Lambda \right)
  \right] + {\cal O} (g^5).
\end{align}

The condition \eqref{PV-subgauge} is satisfied by the field in
\eq{eq:A_LC_NLOlong} only if
\begin{align}
  \label{eq:diff_true}
  {\nabla}^2_\perp C_2 ({\vec x}_\perp, b_1, b_2) & = - \frac{i}{8}
  \frac{g^4}{2 \pi} \, [t^a (t^a)_1 , t^b (t^b)_2 ] \, \left[ \delta^2
    \!  \left( {\vec x}_\perp - {\vec b}_{2\perp} \right) - \delta^2
    \!  \left( {\vec x}_\perp - {\vec b}_{1\perp} \right) \right] \,
  \ln \left(|{\vec b}_{1\perp} - {\vec b}_{2\perp}| \, \Lambda
  \right).
\end{align}
The solution of \eq{eq:diff_true} is
\begin{align}
  \label{eq:C2sol}
  C_2 ({\vec x}_\perp, b_1, b_2) = - \frac{i \, g^4}{8 \, (2\pi)^2} \,
  [t^a (t^a)_1 , t^b (t^b)_2 ] \, \ln \left(\frac{|{\vec x}_\perp -
      {\vec b}_{2\perp}|}{|{\vec x}_{\perp} - {\vec b}_{1\perp}|}
  \right) \, \ln \left(|{\vec b}_{1\perp} - {\vec b}_{2\perp}| \,
    \Lambda \right),
\end{align}
where we put integration constants to zero and required that $C_2$ is
at most finite as $x_\perp \to \infty$ such that ${\vec A}_\perp^{LC}
\to 0$ when $x_\perp \to \infty$, which was our assumption throughout
the paper. Substituting \eq{eq:C2sol} into \eq{eq:A_LC_NLOlong} we
obtain our final result for the gluon field in light-cone gauge,
\begin{align}
  \label{eq:A_LC_NLOfinal}
  {\vec A}_\perp^{LC} (x^-, {\vec x}_\perp) & = \frac{g}{4 \pi} \, t^a
  (t^a)_1 \, \mbox{Sign} (x^- - b_{1}^- ) \, \frac{{\vec x}_\perp -
    {\vec b}_{1\perp}}{|{\vec x}_\perp - {\vec b}_{1\perp}|^2} +
  \frac{g}{4 \pi} \, t^a (t^a)_2 \, \mbox{Sign} (x^- - b_{2}^- ) \,
  \frac{{\vec x}_\perp - {\vec b}_{2\perp}}{|{\vec x}_\perp - {\vec
      b}_{2\perp}|^2} \notag \\ & - \frac{i}{8} \frac{g^3}{(2 \pi)^2}
  \, [t^a (t^a)_1 , t^b (t^b)_2 ] \, \mbox{Sign} (b_2^- - b_{1}^- ) \,
  \left[ \mbox{Sign} (x^- - b_{2}^- ) + \mbox{Sign} (x^- - b_{1}^- )
  \right] \notag \\ & \times \left[ \frac{{\vec x}_\perp - {\vec
        b}_{1\perp}}{|{\vec x}_\perp - {\vec b}_{1\perp}|^2} \, \ln
    \left(|{\vec x}_\perp - {\vec b}_{2\perp}| \, \Lambda \right) +
    \frac{{\vec x}_\perp - {\vec b}_{2\perp}}{|{\vec x}_\perp - {\vec
        b}_{2\perp}|^2} \, \ln \left(|{\vec x}_\perp - {\vec
        b}_{1\perp}| \, \Lambda \right) \right] \notag \\ & +
  \frac{i}{8} \frac{g^3}{(2 \pi)^2} \, [t^a (t^a)_1 , t^b (t^b)_2 ] \,
  \left[ \frac{{\vec x}_\perp - {\vec b}_{2\perp}}{|{\vec x}_\perp -
      {\vec b}_{2\perp}|^2} - \frac{{\vec x}_\perp - {\vec
        b}_{1\perp}}{|{\vec x}_\perp - {\vec b}_{1\perp}|^2} \right]
  \, \ln \left(|{\vec b}_{1\perp} - {\vec b}_{2\perp}| \, \Lambda
  \right) \notag \\ & - \frac{i}{8} \frac{g^3}{(2 \pi)^2} \, [t^a
  (t^a)_1 , t^b (t^b)_2 ] \, \mbox{Sign} (x^- - b_{1}^- ) \,
  \mbox{Sign} (x^- - b_{2}^- ) \notag \\ & \times \left[ \frac{{\vec
        x}_\perp - {\vec b}_{2\perp}}{|{\vec x}_\perp - {\vec
        b}_{2\perp}|^2} \, \ln \left(|{\vec x}_\perp - {\vec
        b}_{1\perp}| \, \Lambda \right) - \frac{{\vec x}_\perp - {\vec
        b}_{1\perp}}{|{\vec x}_\perp - {\vec b}_{1\perp}|^2} \, \ln
    \left(|{\vec x}_\perp - {\vec b}_{2\perp}| \, \Lambda \right)
  \right] + {\cal O} (g^5).
\end{align}

It is important to stress that imposing a stronger sub-gauge
condition \eqref{PVwrong} onto the field of \eq{eq:A_LC_NLOlong} would
lead to
\begin{align}
  \label{eq:diff2}
  {\vec \nabla}_\perp C_2 ({\vec x}_\perp, b_1, b_2) = - \frac{i}{8}
  \frac{g^4}{(2 \pi)^2} \, [t^a (t^a)_1 , t^b (t^b)_2 ] \left[
    \frac{{\vec x}_\perp - {\vec b}_{2\perp}}{|{\vec x}_\perp - {\vec
        b}_{2\perp}|^2} \, \ln \left(|{\vec x}_\perp - {\vec
        b}_{1\perp}| \, \Lambda \right) - \frac{{\vec x}_\perp - {\vec
        b}_{1\perp}}{|{\vec x}_\perp - {\vec b}_{1\perp}|^2} \, \ln
    \left(|{\vec x}_\perp - {\vec b}_{2\perp}| \, \Lambda \right)
  \right].
\end{align}
However, \eq{eq:diff2} for $C_2$ has no solution. The easiest way to
see it is to act on both sides with ${\vec \nabla}_\perp \times$,
\begin{align}
  \label{eq:diff3}
  0 = {\vec \nabla}_\perp \times {\vec \nabla}_\perp C_2 ({\vec
    x}_\perp, b_1, b_2) \neq - \frac{i}{4} \frac{g^4}{(2 \pi)^2} \,
  [t^a (t^a)_1 , t^b (t^b)_2 ] \, \frac{{\vec x}_\perp - {\vec
      b}_{1\perp}}{|{\vec x}_\perp - {\vec b}_{1\perp}|^2} \times
  \frac{{\vec x}_\perp - {\vec b}_{2\perp}}{|{\vec x}_\perp - {\vec
      b}_{2\perp}|^2},
\end{align}
obtaining a contradiction. (Here ${\vec \nabla}_\perp \times {\vec
  a}_\perp \equiv \pd_x \, a_y - \pd_y \, a_x$.)

We conclude that one can not {\sl always} satisfy the condition
\eqref{PVwrong} in a Yang-Mills theory: we have just constructed a
counter-example. Therefore \eq{PVwrong} is not a proper sub-gauge
condition of the light-cone gauge, which did not follow from our
discussion in Sec.~\ref{sec:PV}. At the same time the condition
\eqref{PV-subgauge} appears to have passed this non-Abelian classical
field test leading to the gluon field
\eqref{eq:A_LC_NLOfinal}.\footnote{One may argue that the condition
  \eqref{PVwrong} is actually two conditions, due to its (two-)vector
  nature, and it may over-constrain the system, whereas the condition
  \eqref{PV-subgauge} is only one condition, being a scalar under
  rotations in the transverse plane. However, presently we can not
  construct a proof of this conjecture in the general case.}


\subsection{Diagrammatic Calculation}

To better understand what using the PV prescription for the
propagators \eqref{propagator-PV} entails in the actual diagrammatic
calculations, let us now try to construct the gluon field of two
ultrarelativistic color charges using Feynman diagrams. 

\begin{figure}[ht]
\begin{center}
\includegraphics[width=0.6 \textwidth]{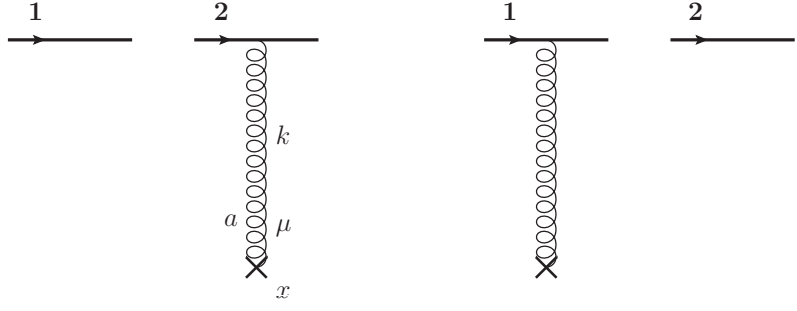} 
\caption{Diagrammatic representation of the gluon field of two quarks
  at the order $g$.}
\label{fig:LO}
\end{center}
\end{figure}

We start with the order-$g$ gluon field of two quarks in the
light-cone gauge depicted in \fig{fig:LO}. A straightforward
calculation (using PV regularization of the light-cone singularities)
yields
\begin{align}
  \label{eq:LOfield}
  {\vec A}^{LC}_\perp (x^-, {\vec x}_\perp) = \, & t^a \, \int
  \frac{d^2 k_\perp \, d k^+}{(2 \pi)^3} \, e^{- i k^+ (x^- - b_1^-) +
    i {\vec k}_\perp \cdot ({\vec x}_\perp - {\vec b}_{1\perp})} \, g
  (t^a)_1 \, \frac{k_\perp^\mu}{k_\perp^2} \, \mbox{PV} \left\{
    \frac{1}{k^+} \right\} + (1 \to 2) \notag \\ = \, & \frac{g}{4
    \pi} t^a \, (t^a)_1 \, \mbox{Sign} (x^- - b_1^-) \, \frac{{\vec
      x}_\perp - {\vec b}_{1\perp}}{|{\vec x}_\perp - {\vec
      b}_{1\perp}|^2} + \frac{g}{4 \pi} t^a \, (t^a)_2 \, \mbox{Sign}
  (x^- - b_2^-) \, \frac{{\vec x}_\perp - {\vec b}_{2\perp}}{|{\vec
      x}_\perp - {\vec b}_{2\perp}|^2}
\end{align}
in agreement with \eq{eq:A_LC_LO2}.

\begin{figure}[ht]
\begin{center}
\includegraphics[width=0.8 \textwidth]{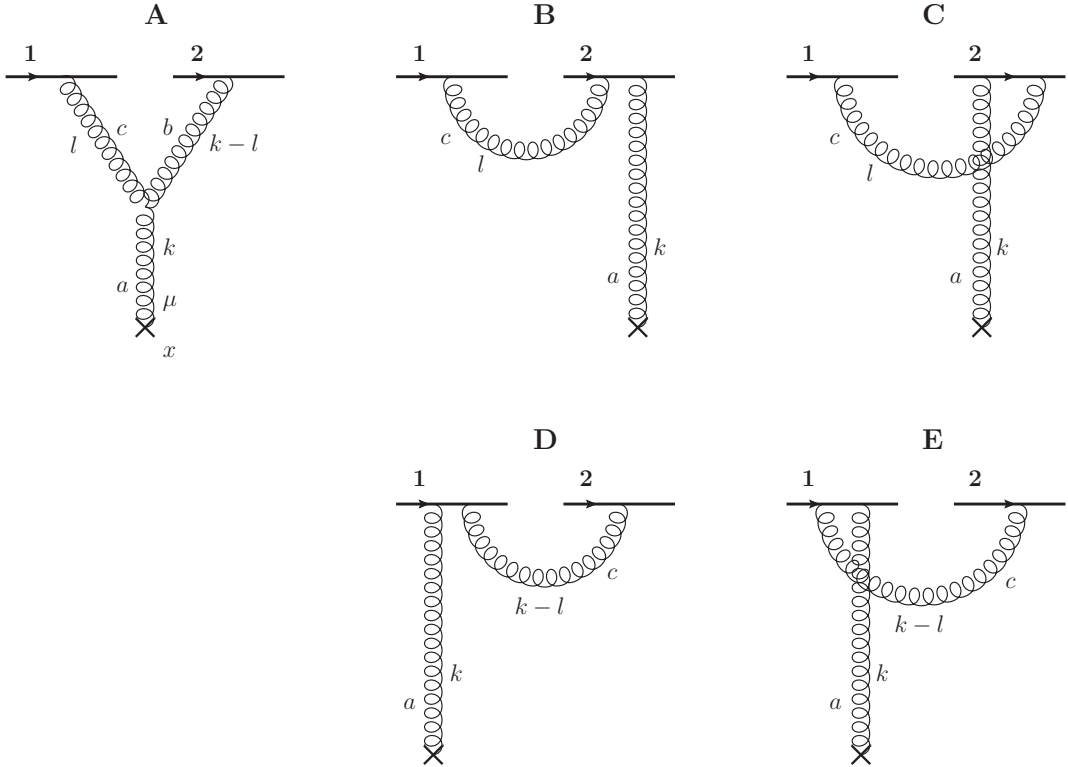} 
\caption{Diagrammatic representation of the classical gluon field of
  two quarks at the order $g^3$.}
\label{fig:NLO}
\end{center}
\end{figure}

Now let us explore the next-to-lowest order. Diagrams contributing to
the order-$g^3$ classical field are shown in \fig{fig:NLO}
(cf. \cite{Kovchegov:1997pc}). A straightforward but a little more
tedious calculation yields (in $k^+, {\vec k}_\perp$ momentum space)
\begin{subequations}
  \begin{align}
    & A = i \, g^3 \, f^{abc} \, (t^b)_2 \, (t^c)_1 \frac{1}{k_\perp^2
      \, l_\perp^2 \, ({\vec k}_\perp - {\vec l}_\perp)^2} \, \left[
      \frac{- k_\perp^2 \, l_\perp^\mu + {\vec k}_\perp \cdot {\vec
          l}_\perp \, k_\perp^\mu}{l^+ \, (k^+ - l^+)} + \frac{{\vec
          l}_\perp \cdot ({\vec k}_\perp - {\vec l}_\perp) \,
        k_\perp^\mu \, (k^+ - 2 l^+)}{k^+ \, l^+ \, (k^+ - l^+)}
    \right], \\
    & B+C = i \, g^3 \, f^{abc} \, (t^b)_2 \, (t^c)_1
    \frac{k_\perp^\mu}{k_\perp^2 \, l_\perp^2} \, \frac{1}{k^+ \, l^+}, \\
    & D+E = - i \, g^3 \, f^{abc} \, (t^b)_2 \, (t^c)_1
    \frac{k_\perp^\mu}{k_\perp^2 \, ({\vec k}_\perp - {\vec
        l}_\perp)^2} \, \frac{1}{k^+ \, (k^+ - l^+)}.
  \end{align}
\end{subequations}

The light-cone gauge gluon field due to the sum of the diagrams $A$
through $E$ is
\begin{align}
  \label{eq:NLOfield}
  {\vec A}^{LC}_\perp (x^-, {\vec x}_\perp) = \, t^a & \int \frac{d^2
    k_\perp \, d k^+}{(2 \pi)^3} \, \frac{d^2 l_\perp \, d l^+}{(2
    \pi)^3} \, e^{- i k^+ (x^- - b_2^-) -i l^+ (b_2^- - b_1^-) + i
    {\vec k}_\perp \cdot ({\vec x}_\perp - {\vec b}_{2\perp}) + i
    {\vec l}_\perp \cdot ({\vec b}_{2\perp} - {\vec b}_{1\perp})} \, i
  \, g^3 \, f^{abc} \, (t^b)_2 \, (t^c)_1 \notag \\ \times
  \frac{1}{k_\perp^2 \, l_\perp^2 \, ({\vec k}_\perp - {\vec
      l}_\perp)^2} & \left[ \frac{- k_\perp^2 \, l_\perp^\mu + {\vec
        k}_\perp \cdot {\vec l}_\perp \, k_\perp^\mu}{l^+ \, (k^+ -
      l^+)} + \frac{{\vec l}_\perp \cdot ({\vec k}_\perp - {\vec
        l}_\perp) \, k_\perp^\mu \, (k^+ - 2 l^+)}{k^+ \, l^+ \, (k^+
      - l^+)} + \frac{({\vec k}_\perp - {\vec l}_\perp)^2 \,
      k_\perp^\mu}{k^+ \, l^+} - \frac{l_\perp^2 \, k_\perp^\mu}{k^+
      \, (k^+ - l^+)} \right].
\end{align}
The regularization of all light-cone singularities in \eq{eq:NLOfield}
is (implicitly) PV. All Fourier transforms in \eq{eq:NLOfield} are
well-defined, except for the second term in the square
brackets. There, the integral over $k^+$ and $l^+$ contains pinched
poles. If we were regulating all the light-cone singularities by using
the PV prescription {\it ad hoc}, with different $i \epsilon$'s for
different poles, this integral would have been ill-defined, being
strongly dependent on the order in which different $\epsilon$'s are
sent to zero. However, since all our light-cone propagators
\eqref{propagator-PV} follow from the same generating functional
\eqref{prop-genf}, they all come with the same $i \epsilon$'s. Hence,
as a result of our calculation in Sec.~\ref{sec:PV} we have a specific
prescription for the pinched-pole integral in question: use the same
$i \epsilon$'s for all the light-cone poles in all the gluon
propagators involved. Note that this prescription was used before in
the diagrammatic calculation of next-to-leading order
Dokshitzer--Gribov--Lipatov--Altarelli--Parisi (DGLAP)
\cite{Dokshitzer:1977sg,Gribov:1972ri,Altarelli:1977zs} anomalous
dimensions in \cite{Curci:1980uw}: here we hope to have provided a
justification for this prescription.

To illustrate our prescription explicitly, let us first perform all
the Fourier transforms in \eq{eq:NLOfield} except for the pinched-pole
integral. We get
\begin{align}
  \label{eq:NLOfield2}
  {\vec A}^{LC}_\perp (x^-, {\vec x}_\perp) = \, & - \frac{g^3}{4(2
    \pi)^2} \, t^a \, f^{abc} \, (t^b)_2 \, (t^c)_1 \left\{
    \frac{1}{2} \mbox{Sign} (x^- - b_1^-) \, \mbox{Sign} (x^- - b_2^-)
    \left[ \frac{{\vec x}_\perp - {\vec b}_{2\perp}}{|{\vec x}_\perp -
        {\vec b}_{2\perp}|^2} \, \ln \left(\frac{|{\vec x}_\perp -
          {\vec b}_{1\perp}|}{|{\vec b}_{1\perp} - {\vec b}_{2\perp}|}
      \right) \right. \right. \notag \\ & \left. - \frac{{\vec
        x}_\perp - {\vec b}_{1\perp}}{|{\vec x}_\perp - {\vec
        b}_{1\perp}|^2} \, \ln \left( \frac{|{\vec x}_\perp - {\vec
          b}_{2\perp}|}{|{\vec b}_{1\perp} - {\vec b}_{2\perp}|}
    \right) \right] \notag \\ & + \, \mbox{Sign} (b_2^- - b_1^-) \,
  \mbox{Sign} (x^- - b_2^-) \, \frac{{\vec x}_\perp - {\vec
      b}_{2\perp}}{|{\vec x}_\perp - {\vec b}_{2\perp}|^2} \, \ln
  \left(|{\vec b}_{1\perp} - {\vec b}_{2\perp}| \, \Lambda \right)
  \notag \\ & + \, \mbox{Sign} (b_2^- - b_1^-) \, \mbox{Sign} (x^- -
  b_1^-) \, \frac{{\vec x}_\perp - {\vec b}_{1\perp}}{|{\vec x}_\perp
    - {\vec b}_{1\perp}|^2} \, \ln \left(|{\vec b}_{1\perp} - {\vec
      b}_{2\perp}| \, \Lambda \right) \notag \\ & - 4 \int \frac{d k^+
    \, d l^+}{(2 \pi)^2} \, e^{- i k^+ (x^- - b_2^-) -i l^+ (b_2^- -
    b_1^-)} \frac{k^+ - 2 l^+}{k^+ \, l^+ \, (k^+ - l^+)} \frac{1}{2}
  \left[ \frac{{\vec x}_\perp - {\vec b}_{2\perp}}{|{\vec x}_\perp -
      {\vec b}_{2\perp}|^2} \, \ln \left(\frac{|{\vec x}_\perp - {\vec
          b}_{1\perp}|}{|{\vec b}_{1\perp} - {\vec b}_{2\perp}|}
    \right) \right. \notag \\ & \left. \left. + \frac{{\vec x}_\perp -
        {\vec b}_{1\perp}}{|{\vec x}_\perp - {\vec b}_{1\perp}|^2} \,
      \ln \left(\frac{|{\vec x}_\perp - {\vec b}_{2\perp}|}{|{\vec
            b}_{1\perp} - {\vec b}_{2\perp}|} \right) \right]
  \right\}.
\end{align}
Using the same $i \epsilon$'s to regulate all the poles in the pinched
integral (similar to \cite{Curci:1980uw}) while using the PV
prescription we get
\begin{align}
  \label{eq:v1}
  - 4 \int \frac{d k^+ \, d l^+}{(2 \pi)^2} \, e^{- i k^+ (x^- -
    b_2^-) -i l^+ (b_2^- - b_1^-)} \frac{k^+ - 2 l^+}{k^+ \, l^+ \,
    (k^+ - l^+)} & = - 4 \int \frac{d k^+ \, d l^+}{(2 \pi)^2} \, e^{-
    i k^+ (x^- - b_2^-) -i l^+ (b_2^- - b_1^-)} \frac{(k^+ - l^+) -
    l^+}{k^+ \, l^+ \, (k^+ - l^+)} \notag \\ & \, = \mbox{Sign}
  (b_2^- - b_1^-) \left[ \mbox{Sign} (x^- - b_1^-) + \mbox{Sign} (x^-
    - b_2^-) \right].
\end{align}
As one can show, using \eq{eq:v1} in \eq{eq:NLOfield2} gives
\eq{eq:A_LC_NLOfinal}. (An identity
\begin{align}
  \label{eq:identity}
  1 = \mbox{Sign} (x^- - b_1^-) \, \mbox{Sign} (x^- - b_2^-) +
  \mbox{Sign} (b_2^- - b_1^-) \left[ \mbox{Sign} (x^- - b_1^-) -
    \mbox{Sign} (x^- - b_2^-) \right]
\end{align}
comes in handy.) Hence the same-$i \epsilon$'s prescription is a
diagrammatic equivalent of using the sub-gauge condition
\eqref{PV-subgauge} in the classical field calculations.


\section{Summary}
\label{sec:conclusions}

In this paper we have studied the question of whether the ambiguity
associated with the regularization of the poles of the light-cone
gauge gluon propagator can be eliminated by fixing the residual gauge
freedom using a sub-gauge condition. We saw that this is indeed the
case for the $\theta$-function sub-gauges and for the PV sub-gauge. In
the process we have elucidated the proper sub-gauge condition for the
PV sub-gauge. Our main results for the propagators and for the
sub-gauge conditions are given in (and above)
Eqs.~\eqref{propagator-1} and \eqref{propagator-2} for the
$\theta$-function sub-gauges and by Eqs.~\eqref{propagator-PV} and
\eqref{PV-subgauge} for the PV sub-gauge.

We have also shown that one can construct the classical gluon field of
a single ultrarelativistic nucleus in the PV sub-gauge: our
perturbative calculation for the two ultrarelativistic color charges
resulted in \eq{eq:A_LC_NLOfinal} for the gluon field. Moreover, it
appears that we have constructed a justification for the same-$i
\epsilon$'s prescription for dealing with the light-cone gauge gluon
propagator poles in the PV sub-gauge.


\section*{Acknowledgments}

The authors are grateful to Ian Balitsky and Al Mueller for
encouraging discussions. We would also like to thank Ian Balitsky for
correspondence on the issue in which he stressed the importance of the
surface terms in the functional integration. This material is based
upon work supported by the U.S. Department of Energy, Office of
Science, Office of Nuclear Physics under Award
Number DE-SC0004286. \\


\appendix

\section{}
\renewcommand{\theequation}{A\arabic{equation}}
  \setcounter{equation}{0}
\label{B}

In Section~\ref{sec:theta_ftn} we have imposed \eqref{subgauge} as the
sub-gauge condition requiring the transverse divergence of the gauge
field to be zero at a generic point $x^-=\sigma$. An alternative
sub-gauge condition is for the four-divergence to be zero at a generic
point $x^-=\sigma$:
\begin{align}
\partial_\mu A^\mu(x^-=\sigma) =0 ~.
\label{4divsubg}
\end{align}
Here we will show that the sub-gauge choice (\ref{4divsubg}) is not
suitable for specifying the prescription of the $k^+ =0$ pole of the
light-cone gauge gluon propagator.

The propagator with sub-gauge condition (\ref{4divsubg}) should
satisfy the following differential equation (cf. \eq{greenf-eq})
\begin{align}\label{eq:B}
  \left[ \partial^2 g_{\mu\rho} - \partial_\mu\partial_\rho - {1\over
      \xi_1}\tilden_\mu \tilden_\rho + {1\over
      \xi_2}\partial_\mu\delta(x^- -\sigma)\partial_\rho \right] \,
  {D}^{\rho\nu}(x, y) = i \, \delta_\mu^\nu\,\delta^{(4)}(x-y)~.
\end{align}
Note that $\partial_\mu$ to the left of the delta-function in
\eq{eq:B} acts on everything to its right.

Projecting \eq{eq:B} onto $\eta^\mu \, {\tilde \eta}_\nu$ we get
(cf. Eq.~(75) in \cite{Das:2004qk})
\begin{align}\label{proj}
  \pd^+ \, \left[ \left( 1 - {1\over \xi_2} \, \delta(x^--\sigma)
    \right) \, \partial_\rho \, D^{\rho -} (x,y) \right] + \pd^2
  D^{+-} (x,y) = - 2 i \, \delta^{(4)}(x-y).
\end{align}
This equation has no solution for finite $\sigma$. To see this one can
integrate both sides over $x^-$ in an infinitesimal interval near
$x^-=\sigma$: the contribution of the $\delta$-function term on the
left-hand side of \eq{proj} to such an integral is ill-defined, as it
contains $\delta(x^--\sigma) \Big|_{x^- = \sigma - \epsilon}^{x^- =
  \sigma + \epsilon}$. (If we assume that $\delta(x^--\sigma)
\Big|_{x^- = \sigma - \epsilon}^{x^- = \sigma + \epsilon} =0$ we can
simply drop the second term in \eq{proj}: however, we are not going to
get a regularization of the $k^+=0$ poles this way.) The only way to
avoid this ambiguity is to require that $\partial_\rho \, D^{\rho -}
(x^- = \sigma,y) =0$, which may only be true for $\sigma = \pm \infty$
(see a similar discussion near \eq{cond-D} in the main text). This
would result in the propagators \eqref{propagator-1} or
\eqref{propagator-2} corresponding to $\sigma = \pm \infty$. However,
for $\sigma = \pm \infty$ we saw in \eq{x-boundary} that the sub-gauge
condition \eqref{4divsubg} does not give zero, and hence does not
work.

To summarize, we see that for finite $\sigma$ no solution of \eq{eq:B}
exists, while for $\sigma = \pm \infty$ the solution does not satisfy
the boundary condition in \eq{x-boundary}. From this we conclude that
$\partial_\mu A^\mu(x^-=\sigma) = 0$ is not a suitable sub-gauge
condition for the light-cone gauge in the functional integral
formalism.

For pedagogical reasons let us arrive at the same conclusion using a
slightly different technique. It is convenient to introduce the
following two linearly independent tensors structures orthogonal to
$\eta^\mu$,
\begin{subequations}\label{tensorab}
\begin{align}
  & a^{\mu\nu} \equiv g^{\mu\nu} - {\partial^\mu \tilden^\nu
    +\partial^\nu \tilden^\mu\over \tilden\cdot\partial} +
  {\partial^2\tilden^\mu\tilden^\nu \over (\tilden\cdot\partial)^2} -
  {\xi_1\partial^2\partial^\mu\partial^\nu \over
    (\tilden\cdot\partial)^2},
  \\
  & b^{\mu\nu} \equiv - {\partial^2\over
    (\tilden\cdot\partial)^2}\tilden^\mu\tilden^\nu ~,
\end{align}
\end{subequations}
so that we can decompose ${D}^{\mu\nu}(x,y)$ (with the $A^+ =0$ gauge
condition imposed) as
\begin{align} 
{D}^{\mu\nu}(x,y) = a^{\mu\nu} \, a(x,y) + b^{\mu\nu} \,  b(x,y)
\label{decoab}
\end{align}
with functions $a(x,y)$ and $b(x,y)$ to be determined.

Using (\ref{decoab}) in \eqref{eq:B} we have
\begin{align}
  & \left[\delta^\nu_\mu - {\partial_\mu\tilden^\nu \over \partial^+}
    +\partial^2{\tilden_\mu\tilden^\nu\over \partial^{+2}} -
    {\xi_1\over \xi_2}\partial_\mu\delta(x^--\sigma) \partial^\nu
    {\partial^2\over \partial^{+2}}\right]\partial^2 a(x,y)
  \nonumber\\
  & + \left[-{\partial^2\tilden_\mu\tilden^\nu \over \partial^{+2}} +
    {\partial_\mu\tilden^\nu\over \partial^+} - {1\over
      \xi_2}\partial_\mu\delta(x^--\sigma){\tilden^\nu\over \partial^+}\right]
\partial^2 b(x,y) = i \, \delta^{(4)}(x-y) \, \delta^\nu_\mu~.
\label{decomp-eq}
\end{align}
Projecting \eq{decomp-eq} onto $\eta^\mu \, {\tilde \eta}_\nu$ again
we get (cf. \eq{proj})
\begin{align}
  \Big[1-{1\over
    \xi_2}\partial^+\delta(x^--\sigma){1\over \partial^+}\Big]\partial^2
  b(x,y) = i \, \delta^{(4)}(x-y).
\label{projn}
\end{align}
(Note that we have set $\xi_1$ to zero because at this point the
light-cone gauge has already been employed.) 

Just like \eq{proj}, equation~(\ref{projn}) does not provide any
prescription for the $k^+=0$ pole for any finite $\sigma$.  For
$\sigma = \pm \infty$ we have already seen that sub-gauge condition
(\ref{4divsubg}) is not compatible with the path integral
formalism. From this analysis we again conclude that the sub-gauge
$\partial_\mu A^\mu(x^-=\sigma) = 0$ is not a suitable sub-gauge of
the light-cone gauge in the functional integral formalism.


\section{}
\renewcommand{\theequation}{B\arabic{equation}}
  \setcounter{equation}{0}
\label{A}

In this Appendix we provide details of the calculation carried out in
\eq{x-boundary} (as well as those in Eqs.~\eqref{x-boundaryPV}) and
\eqref{x-boundaryML}). More specifically, in the transition from the
second to the third line of \eq{x-boundary} we neglected the
contributions of the $k^2 =0$ Feynman pole. To justify this let us
consider the Fourier transform of the terms in the square brackets of
the second line of \eq{x-boundary}. The first term is not affected by
the $k^+$ prescription and it is zero at $x^-=\pm \infty$:
\begin{align}
  \int \frac{d^4 k}{(2 \pi)^4} \, e^{-ik\cdot(x-y)}{k^-\over
    k^2+i\epsilon} \, A^\mu (x) \, \Bigg|_{x^-=- \infty}^{x^- = +
    \infty} = {1\over 2\pi^2}{x^- - y^- \over [(x-y)^2-i\epsilon]^2}
  \, A^\mu (x) \, \Big|_{x^-=- \infty}^{x^- = + \infty} = 0.
\label{FT-nosubg-term}
\end{align}
In arriving at zero on the right-hand side of \eqref{FT-nosubg-term}
we assume that $A^\mu (x)/x^- \to 0$ as $x^- \to \infty$, that is that
$A^\mu (x)$ grows slower than $|x^-|$ as $x^- \to \infty$. Note that
the expression in \eq{FT-nosubg-term} is zero at each limiting point,
$x^- = + \infty$ and $x^- = - \infty$, separately.

To understand the $x^-$-dependence of the second terms in the square
brackets of the second line of \eq{x-boundary}, note that $k \cdot
A(x) = k^+ \, A^- (x) - {\vec k}_\perp \cdot {\vec A}_\perp (x)$ in
$A^+ =0$ light-cone gauge. (Once again we assume that the $\xi_1 \to
0$ limit is taken in \eq{gen-funct} enforcing the gauge condition.)
The $k^+ \, A^- (x)$ term vanishes due to \eq{FT-nosubg-term} along
with
\begin{align}
  \label{eq:FT}
  \int \frac{d^4 k}{(2 \pi)^4} \, {k^\mu_\perp \over k^2+i\epsilon} \,
  e^{-ik\cdot (x-y)} = {(x-y)^\mu_\perp\over
    2\pi^2[(x-y)^2-i\epsilon]^2}.
\end{align}

To find the contribution of the ${\vec k}_\perp \cdot {\vec A}_\perp
(x)$ we use the following integral
\begin{align}
  & \int \frac{d^4 k}{(2 \pi)^4} {k^i_\perp \, e^{-ik\cdot(x-y)}\over
    (k^2+i\epsilon)(k^+ + i\epsilon)} = -{(x-y)^i_\perp\over
    2\pi(x-y)^2_\perp} \, \theta(x^- - y^-) \, \delta(x^+ - y^+)
  \nonumber\\
  & + {i\,(x-y)^i_\perp\over 2\pi^2(x-y)^2_\perp} \, (x^- - y^-)
  \left[ \frac{1}{(x-y)^2 - i \epsilon} - \frac{1}{2 (x^+-y^+) (x^-
      -y^-) - i \epsilon} \right] ~ .
\label{FT-subg-term}
\end{align}
Note that the $k^2 =0$ pole gives the second term on the right-hand
side, which vanishes as $x^- \to \infty$. Rewriting $k^- \eta^\mu +
k_\perp^\mu \to i \partial^- \eta^\mu + i \partial_\perp^\mu$ (all
derivatives are with respect to $x$) and noticing that applying
derivatives to the second term on the right-hand side of
\eq{FT-subg-term} would still leave it vanishing at $x^- \to \infty$,
we complete the justification of neglecting the contributions of the
$k^2 =0$ pole in going from the second to the third line of
\eq{x-boundary}. (Once again we have to assume that $A^\mu (x)/x^- \to
0$ as $x^- \to \infty$.) The first term in \eq{FT-subg-term} does not
vanish for $x^- \to + \infty$: this term is due to picking up the $k^+
=0$ pole and is the one giving us the third line of \eq{x-boundary}.

The conclusion reached here about the $k^2 =0$ pole contribution
vanishing at $x^- \to \infty$ is independent of the regularization of
the $k^+ =0$ pole and thus applies to PV and ML sub-gauges as well.



\providecommand{\href}[2]{#2}\begingroup\raggedright\endgroup

\end{document}